\documentclass[twocolumn,showpacs,amsmath,aps]{revtex4}
\usepackage{graphicx,color}
\usepackage{bm}
\usepackage[hypertex]{hyperref}

\newcommand{\be}{\begin{equation}}
\newcommand{\ee}{\end{equation}}
\newcommand{\bea}{\begin{eqnarray}}
\newcommand{\eea}{\end{eqnarray}}
\newcommand{\bsube}{\begin{subequations}}
\newcommand{\esube}{\end{subequations}}

\newcommand{\Eq}[1]{Eq.\,(\ref{#1})}
\newcommand{\Eqs}[1]{Eqs.\,(\ref{#1})}

\newcommand{\dg}{\dagger}
\newcommand{\la}{\langle}
\newcommand{\ra}{\rangle}

\newcommand{\ep}{\epsilon}

\newcommand{\al}{\alpha}

\newcommand{\ti}{\tilde}
\newcommand{\nl}{\nonumber \\}
\newcommand{\nla}{\nl&\quad}
\newcommand{\up}{\uparrow}
\newcommand{\down}{\downarrow}
\newcommand{\mb}{\mbox}
\newcommand{\alp}{\alpha}

\newcommand{\gam}{\gamma}

\newcommand{\eps}{\epsilon}
\newcommand{\vep}{\varepsilon}

\newcommand{\lam}{\lambda}    % == Cao

\newcommand{\omg}{\omega}

\newcommand{\wit}{\widetilde}

\newcommand{\sgm}{\sigma}
\newcommand{\Gam}{\Gamma}
\newcommand{\bsub}{\begin{subequations}}
\newcommand{\esub}{\end{subequations}}

\newcommand{\re}{\nonumber\\}
\newcommand{\ket}[1]{{\left| #1 \right\rangle }}   % == Cao

\newcommand{\cdg}{c^\dagger}
\newcommand{\ddg}{d^\dagger}

\newcommand{\beqn}{\begin{eqnarray}}
\newcommand{\eeqn}{\end{eqnarray}}
\newcommand{\beq}{\begin{equation}}     % === Cao
\newcommand{\eeq}{\end{equation}}

\newcommand{\fdg}{f^\dagger}

\begin{document}
%\begin{CJK*}{GBK}{Song}

\title{Number-resolved master equation approach
        to quantum measurement \\
        and quantum transport}

\author{Xin-Qi Li}
\email{lixinqi@bnu.edu.cn}
\affiliation{Center for Advanced Quantum Studies and
Department of Physics, Beijing Normal University,
Beijing 100875, China}

\date{\today}

%% \maketitle
\begin{abstract}
In addition to the well-known Landauer-B\"uttiker
scattering theory and the nonequilibrium
Green's function technique for mesoscopic transports,
an alternative (and very useful) scheme
is quantum master equation approach.  
In this article, we review the particle-number ($n$)-resolved
master equation ($n$-ME) approach and its systematic applications
in quantum measurement and quantum transport problems.
The $n$-ME contains rich dynamical information,
allowing efficient study of topics such as shot noise
and full counting statistics analysis.
Moreover, we also review a newly developed master equation
approach (and its $n$-resolved version)
under self-consistent Born approximation.
The application potential of this new approach
is critically examined via its ability to recover
the exact results
for noninteracting systems under arbitrary voltage
and in presence of strong quantum interference,
and the challenging non-equilibrium Kondo effect.  
\end{abstract}


\maketitle

\section{Introduction}

{\flushleft Quantum master equation} 
is typically applied for
the reduced state evolution of an open quantum system, e.g.,
in quantum optics and quantum dissipation studies \cite{Car93,Walls94}.
This formalism is also particularly appropriate
for studying quantum measurements, where
the measured subsystem is the {\it system of interest},
and the apparatus is an {\it environment}.
In this context, in most cases, some internal degrees of freedom
of the apparatus should be retained, which may result in
certain back-action effects.
Moreover, for quantum measurement, extra issues
should be taken into account, such as
the readout characteristics of the measurement and
the stochastic evolution of the measured state
{\it conditioned} on the stochastic results of measurement \cite{WM09}.

An interesting solid-state application of quantum measurement
is to measure charge qubits using a mesoscopic detector,
which can be either a quantum-point-contact (QPC) detector
\cite{Gur97,Win97,Moz02,Gur03} or a single electron transistor
(SET) \cite{Dev00,Sch98,Sch01,Ston02}.
For realistic applications of such measurement, the non-trivial
correlation between the detector and qubit has been
the focus of extensive studies
\cite{Gur97,Gur03,Kor01a,Goa01a}.
For instance, for the qubit-QPC setup, if the energy transfer
between the detector and qubit is ignored,
the qubit may relax to invalid statistical mixture
\cite{Shn02,Sta03,Ave04,Li04}.

Moreover, for this type of measurement, we can construct
a particle-number($n$)-resolved master equation ($n$-ME) scheme.  
That is, by properly clarifying the subspace of the apparatus
states in association with the number of electrons transmitted,
one can obtain \cite{Li05,Li05b}
\begin{align}\label{nME}
\dot{\rho}^{(n)}(t)=&-i\mathcal{L}\rho^{(n)}(t)
-\sum_{j=0,\pm 1} \mathcal{R}_j\, \rho^{(n+j)}(t) .
\end{align}
Here, $\rho^{(n)}$ is the (reduced) qubit state
conditioned on the number of electrons ``$n$" transmitted in the detector.
The Liouvillian $\mathcal{L}$ is the well-known commutator
defined by the system Hamiltonian $H_S$.
The superoperators $\mathcal{R}_j$ are associated with
the tunneling processes in the transport detector,
which have explicit forms, as given in \cite{Li05}.
With the knowledge of $\rho^{(n)}(t)$, one is able to carry out
the various readout characteristics of the measurement,  
by noting that the distribution function of the transmitted electrons
is related with the $n$-conditioned density matrix
as $P(n,t)=\mb{Tr}[\rho^{(n)}(t)]$,
where the trace is over the system states.

For the quantum measurement discussed above,
the detector itself is a transport device.
Hence, the $n$-ME approach, \Eq{nME},
is a natural tool for studying quantum transport
through various mesoscopic (nano-scale) devices.
In this context, however, the quantum coherence and/or
many-body interaction effects may more significantly affect the transport properties and device functionalities.
Simply, the master equation approach
is appropriate for quantum transport 
mainly because we can
regard the central device as the {\it system of interest},
and the transport leads (reservoirs) as the generalized {\it environment}.

Compared to the well-known Landauer-B\"uttiker theory \cite{Dat95}
and the nonequilibrium Green's function formalism \cite{Jau96},
the master equation approach
(especially the $n$-ME formulation \cite{Gur96b,Gur96a})
has been very useful for studying
quantum noise in transport \cite{rev0003,Lev9396,Naz01,But02,
Sam05,Sch0506,Bel05,Bk00,Bel04,Bla04,Jau04,Jau05,Bee06,Li07,Ens06}.
One may note that, beyond the usual (average) current,
current fluctuations in mesoscopic transport can provide
useful information for the relevant mechanisms.
Moreover, a fascinating approach, known as full counting
statistics (FCS) analysis \cite{Lev9396}, can conveniently yield
all the statistical cumulants of the number of transferred charges
\cite{Naz01,But02,Sam05,Sch0506,Bel05,Bk00,Bel04,
Bla04,Jau04,Jau05,Bee06,Li07}.
The FCS has been demonstrated experimentally
for transport through quantum dots \cite{Ens06}.

In essence, the $n$-ME provides an important
distribution function via $P(n,t)=\mb{Tr}[\rho^{(n)}(t)]$,
which contains rich information and allows for convenient calculation
of not only the transport current, but also the noise spectrum
and counting statistics.
For instance, for the latter,
all orders of the cumulants of the transmitted
electrons can be calculated by using
$ e^{-{\cal F}(\chi,t)}=\sum_n P(n,t)e^{i n\chi}$,
where $\chi$ is the {\it counting field}
and ${\cal F}(\chi,t)$ is the cumulant generating function (CGF).

In this article, we briefly review the $n$-ME approach and its
applications in quantum measurement and quantum transport problems.
In Sec.\ II,
for the qubit-QPC setup and general quantum transport system,
we first review the key idea and main procedures
for constructing the $n$-ME formalism, and
then outline the methods of calculating
the measurement/transport current and noise spectrum
(using MacDonald's formula).
Particular attention will be given to decomposition
of the $n$-dependent subspaces of the reservoir states
and the consequences of the {\it closed} circuit nature,
which would significantly affect the reservoir state averages.
In Sec.\ III, we discuss the application of the $n$-ME
to two measurement setups in detail,
i.e., a qubit measured by QPC and SET detectors.
In Sec.\ IV, we further discuss the application of the $n$-ME
to quantum transport by using the double-dot Aharonov-Bohm
(DDAB) interferometer and Majorana fermion (MF) probe as examples.
In Sec.\ V, we review the newly proposed self-consistent
Born approximation based master equation (SCBA-ME) approach
to quantum transport; the SCBA-ME scheme
goes beyond the usual master equation approach under
the standard Born approximation which, for instance,
can recover the exact results
of quantum transport through noninteracting systems
and predict the challenging non-equilibrium Kondo effect
for transport through Anderson impurity (interacting dots).
Finally, in Sec.\ VI  we present our concluding remarks.

\section{General Formalism}

{\flushleft In this section, }
we review the construction of the $n$-ME formalism,
and outline the methods of applying it to calculate
the measurement/transport current and noise spectrum.

\subsection{Number($n$)-Resolved Master Equation}

\subsubsection{Set-up (I): Qubit measurement using QPC detector}

\begin{figure}\label{Fig1}
\begin{center}
\centerline{\includegraphics [scale=0.35] {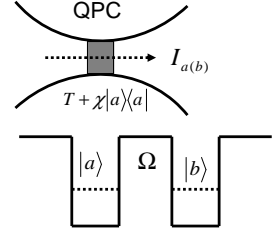}}
\caption{
Schematic for a qubit (coupled quantum dots)
measured by quantum point contact detector. }
\end{center}
\end{figure}

{\flushleft For the sake of generality,}
we formally consider an arbitrary
quantum system measured using a QPC detector.
The whole setup can be described by the Hamiltonian as follows \cite{Li05}:
\begin{subequations} \label{H1}
\begin{eqnarray}
%%\bea\label{H1}
H   &=&  H_0+H'  ,
\\
H_0 &=&  H_{s}   +  \sum_k (\epsilon^L_k c^{\dg}_kc_k+
         \epsilon^R_k d^{\dg}_kd_k) ,
\\
H'  &=& \sum_{k,q} [T_{qk}\{|\psi_s\ra\la \psi_s|\}
         d^{\dg}_q c_k  + \mb{H.c.}] .
\end{eqnarray}
\end{subequations}
In this decomposition, the free part of the total Hamiltonian $H_0$
contains the Hamiltonians of the measured system $H_s$
and the QPC reservoirs (the last two terms).
The Hamiltonian $H'$ describes electron tunneling through the QPC,
e.g., from state $|k\ra$ in the left reservoir to
state $|q\ra$ in the right one,
with a tunneling amplitude of $T_{qk}\{|\psi_s\ra\la \psi_s|\}$
which may depend on the state $|\psi_s\ra$ of the observable.

Regarding the tunneling Hamiltonian $H'$ as a perturbation,
on the basis of the second-order Born expansion, we can derive
a formal equation for the reduced density matrix as \cite{Yan98}
\bea\label{ME-1}
\dot{\rho}(t) = -i {\cal L}\rho(t) - \int^{t}_{0}d\tau \la {\cal L}'(t){\cal G}(t,\tau)
                {\cal L}'(\tau){\cal G}^{\dg}(t,\tau) \ra \rho(t).
\eea
Here, the Liouvillian superoperators are defined as
${\cal L}(\cdots)\equiv [H_s,(\cdots)]$,
${\cal L'}(\cdots)\equiv [H',(\cdots)]$, and
${\cal G}(t,\tau)(\cdots)\equiv G(t,\tau)(\cdots)G^{\dg}(t,\tau)$
with $G(t,\tau)$ the usual propagator (Green's function) associated with $H_s$.
%%%%%%%%%
The reduced density matrix is $\rho(t)=\mb{Tr}_D[\rho_T(t)]$,
resulting from tracing
all the detector degrees of freedom from the entire density matrix.
%%%%
However, for quantum measurement where specific readout information
is to be recorded, the average should be taken
over the unique class of states of the detector
that is being kept track of.

The Hilbert space of the detector can be classified as follows.
First, define the subspace in the absence of electron tunneling
through the detector as ${\cal D}^{(0)}$, which is spanned by the product
of the many-particle states of the two isolated reservoirs, formally denoted
as ${\cal D}^{(0)}\equiv\mb{span}\{|\Psi_L\ra\otimes |\Psi_R\ra \}$.
%%%%
Then, introduce the tunneling operator
$f^{\dg}\sim f^{\dg}_{qk}=d_q^{\dg}c_k$,
and denote the Hilbert subspace corresponding to $n$-electrons
tunneled from the left to the right reservoirs
as ${\cal D}^{(n)}=(f^{\dg})^n {\cal D}^{(0)}$, where $n=1,2,\cdots$.
The entire Hilbert space of the detector is ${\cal D}=\oplus_n{\cal D}^{(n)}$.

With the above classification of the detector states, the average over states
in ${\cal D}$ in \Eq{ME-1} is replaced with states in the subspace
${\cal D}^{(n)}$, leading to a {\it conditional} master equation \cite{Li05}
\bea\label{ME-2}
\dot{\rho}^{(n)}(t) &=& -i {\cal L}\rho^{(n)}(t) - \int^{t}_{0}d\tau
      \mb{Tr}_{D^{(n)}} [{\cal L}'(t){\cal G}(t,\tau)   \nl
      & &  \times {\cal L}'(\tau) {\cal G}^{\dg}(t,\tau) \rho_T(t)] .
\eea
Here, $\rho^{(n)}(t)=\mb{Tr}_{D^{(n)}}[\rho_T(t)]$,
which is the reduced density matrix of the measured system
{\it conditioned} on the number of electrons tunnelled through
the detector until time $t$.
%%%%%
Now, we transform the Liouvillian product in \Eq{ME-2}
to the conventional form:
\bea\label{L-H}
&&  {\cal L}'(t){\cal G}(t,\tau) {\cal L}'(\tau) {\cal G}^{\dg}(t,\tau) \rho_T(t)   \nl
&=& [ H'(t)G(t,\tau)H'(\tau)G^{\dg}(t,\tau)\rho_T(t)  \nl
& &     - G(t,\tau)H'(\tau)G^{\dg}(t,\tau)\rho_T(t)H'(t)] + \mb{H.c.} \nl
&\equiv& [I-II]+\mb{H.c.}
\eea

For simplicity,
we rewrite the interaction Hamiltonian as $H'(t)=QF(t)$.
Here, we have assumed the tunneling amplitude $T_{kq}$ to be real and
independent of the reservoir-state ``$kq$",
and denoted it by $Q$ which depends on the state of the measured system.
The detector fluctuation is described by $F(t)\equiv f(t)+f^{\dg}(t)$,
with $f\equiv\sum_{kq}c^{\dg}_{k}d_q$ and $f^{\dg}\equiv\sum_{kq}d^{\dg}_{q}c_k$.
%%%%
%%%%
Two physical considerations are further made, as follows:
(i)
Instead of the conventional Born approximation for the entire density matrix
$\rho_T(t)\simeq\rho(t)\otimes\rho_D$,
we propose the ansatz $\rho_T(t)\simeq\sum_n\rho^{(n)}(t)\otimes\rho_D^{(n)}$,
where $\rho_D^{(n)}$ is the density operator of the detector reservoirs with
$n$-electrons tunnelled through the detector.
With the ansatz of the density operator, tracing over the subspace
${\cal D}^{(n)}$ yields
\begin{subequations}\label{n-ave}
\bea
\mb{Tr}_{D^{(n)}}[I]&=& \mb{Tr}_D [F(t)F(\tau)\rho_D^{(n)}]  \nl
       & & \times [QG(t,\tau)QG^{\dg}(t,\tau)\rho^{(n)}]  \\
\mb{Tr}_{D^{(n)}}[II]&=& \mb{Tr}_D [f^{\dg}(\tau)\rho_D^{(n-1)}f(t)] \nl
       & & \times [ G(t,\tau)QG^{\dg}(t,\tau)\rho^{(n-1)}Q ]   \nl
       & & + \mb{Tr}_D [f(\tau)\rho_D^{(n+1)}f^{\dg}(t)] \nl
       & & \times [ G(t,\tau)QG^{\dg}(t,\tau)\rho^{(n+1)}Q ] .
\eea
\end{subequations}
Here, we have utilized the orthogonality between states in different subspaces,
which leads to term selection from the entire density operator $\rho_T$.
%%%%
(ii)
Due to the closed nature of the detector circuit, the extra electrons
tunneled into the right reservoir will flow back into the left reservoir
via the external circuit.
In addition, the rapid relaxation processes in the reservoirs will quickly
bring the reservoirs
to the local thermal equilibrium state determined by the chemical potentials.
As a consequence, after the procedure (i.e., the state selection)
as expressed by \Eq{n-ave},
the detector density matrices $\rho_D^{(n)}$ and $\rho_D^{(n\pm 1)}$
in \Eq{n-ave} can be well approximated by $\rho_D^{(0)}$,
i.e., the local thermal equilibrium reservoir state.
%%%%
Under this consideration, the correlation functions become,
$\la f^{\dg}(t)f(\tau)\ra = C^{(+)}(t-\tau)$,
$\la f(t)f^{\dg}(\tau)\ra = C^{(-)}(t-\tau)$,
and $\la F(t)F(\tau)\ra = C(t-\tau)=C^{(+)}(t-\tau)+C^{(-)}(t-\tau)$.
Here, $\la \cdots \ra$ stands for $\mb{Tr}_D [(\cdots)\rho_D^{(0)}]$.

Under the Markovian approximation, the time integral in \Eq{ME-2}
is replaced by $\frac{1}{2}\int^{\infty}_{-\infty}$.
Substituting \Eqs{L-H} and (\ref{n-ave}) into \Eq{ME-2},
we obtain \cite{Li05}
\bea\label{ME-3}
\dot{\rho}^{(n)}
   &=&  -i {\cal L}\rho^{(n)} - \frac{1}{2}
        \left\{  [Q\ti{Q}\rho^{(n)}+\mb{H.c.}] \right. \nl
   & &   - [\ti{Q}^{(-)}\rho^{(n-1)}Q+\mb{H.c.}]   \nl
   & &      \left.  - [\ti{Q}^{(+)}\rho^{(n+1)}Q+\mb{H.c.}] \right\} .
\eea
Here, $\ti{Q}^{(\pm)}=\ti{C}^{(\pm)}({\cal L})Q$,
$\ti{C}^{(\pm)}({\cal L})=\int^{\infty}_{-\infty} dt C^{(\pm)}(t) e^{-i{\cal L}t}$,
and $\ti{Q}=\ti{Q}^{(+)}+\ti{Q}^{(-)}$.
%%%%%%
Under the wide-band approximation for the detector reservoirs,
the spectral function $\ti{C}^{(\pm)}({\cal L})$ can be explicitly
expressed as \cite{Li04}:
$\ti{C}^{(\pm)}({\cal L})
  =  \eta \left[x/(1-e^{-x/T}) \right]_{x=-{\cal L}\mp V}$,
where $\eta=2\pi g_Lg_R$, and $T$ is the temperature.
(Here, and in the following, we use the unit system of $\hbar=e=k_B=1$).
%%%%%%
In \Eq{ME-3}, the terms in $\{\cdots\}$ describe the effect of fluctuation
of forward and backward electron tunneling through the detector
on the measured system.
%%%%%%%
In particular,
the Liouvillian operator ``${\cal L}$" in $\ti{C}^{(\pm)}({\cal L})$
contains the information of energy transfer
between the detector and the measured system, which correlates the energy
(spontaneous) relaxation of the measured system
with the inelastic electron tunneling in the detector.
%%%%%%
At high-voltage limit, formally $V\gg {\cal L}$, the spectral function
$\ti{C}^{(\pm)}({\cal L})\simeq \ti{C}^{(\pm)}(0)$, and \Eq{ME-3} reduce
to the result derived by Gurvitz {\it et al}
\cite{Gur97,Moz02,Gur03,Goa01a}.

\subsubsection{Set-up (II): Quantum transport}

\begin{figure} %\label{fig1}
\includegraphics*[scale=1]{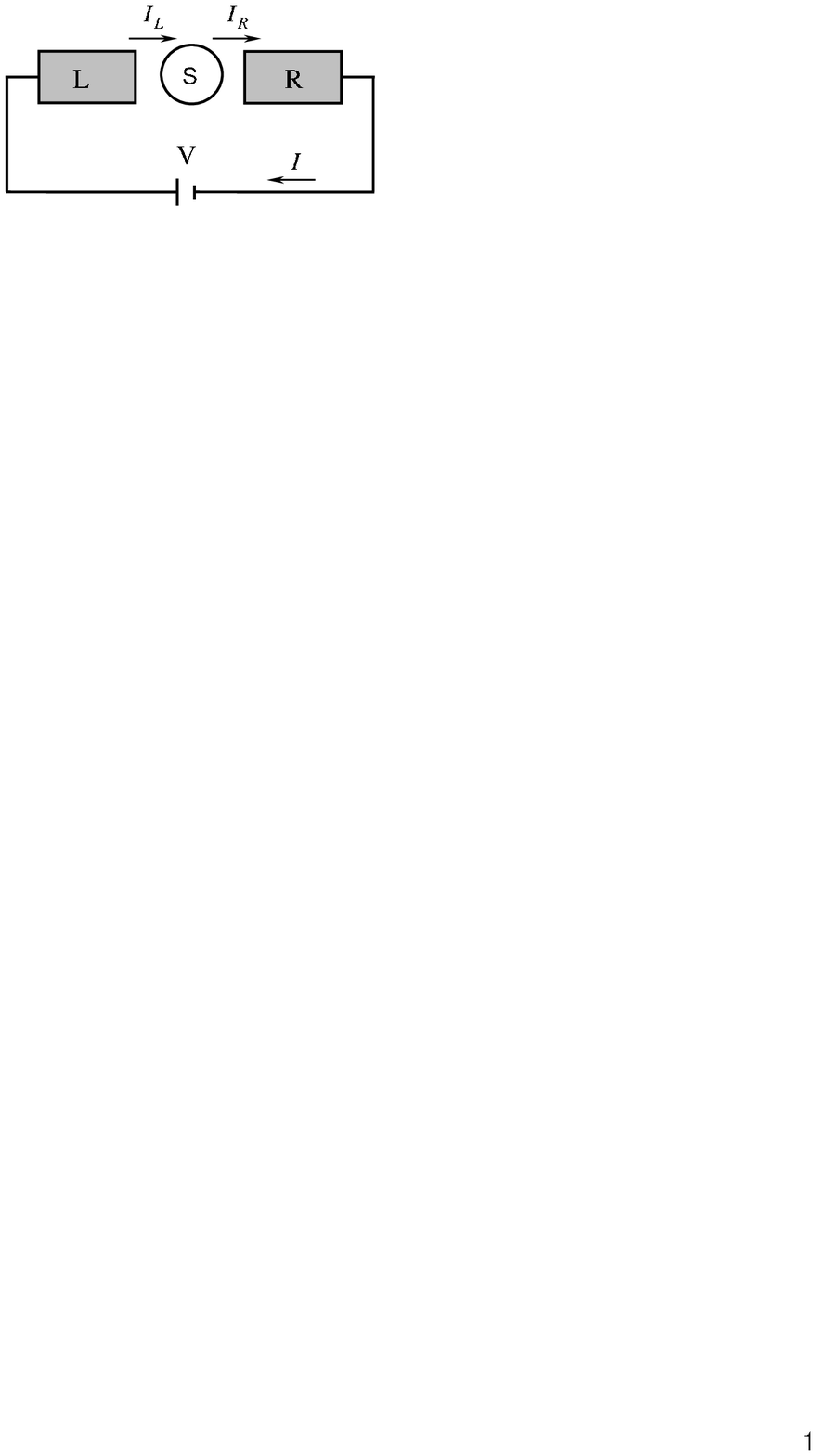}
\caption{Schematic setup for quantum transport
through a mesoscopic system.}
\end{figure}

{\flushleft In general,}
quantum transport, schematically shown in Fig.\ 2,
can be described by the following Hamiltonian:
%%\begin{subequations}
\bea\label{H-ms}
H &=& H_S(a_{\mu}^{\dg},a_{\mu})+\sum_{\al=L,R}\sum_{\mu k}
       \ep_{\al\mu k}d^{\dg}_{\al\mu k}d_{\al\mu k}  \nl
  & &  + \sum_{\al=L,R}\sum_{\mu k}(t_{\al\mu k}a^{\dg}_{\mu}
       d_{\al\mu k}+\rm{H.c.}) .
\eea
%%\end{subequations}
Here, $H_S$ is the central system (device) Hamiltonian,
which can be quite general (e.g., includes many-body interactions).
$a^{\dg}_{\mu}$ ($a_{\mu}$) is the creation (annihilation) operator
of electrons in state $|\mu\ra$,
which indicates both the orbital and spin degrees of freedom.
%%%%
The second and third terms describe, respectively,
the two (left and right) leads (reservoirs)
and the tunneling between them and the central system.
The lead electrons are also attached here
by index ``$\mu$" to characterize their possible
correlation with the {\it system states}. For instance, this
will be the typical situation for spin-dependent transport.

If we introduce
$F_{\mu} = \sum_{\al k} t_{\al\mu k}d_{\al\mu k}
  \equiv f_{L\mu} + f_{R\mu}$
and express the tunneling Hamiltonian as
%%\begin{subequations} \label{Ht}
\begin{eqnarray}\label{Ht}
H' &=& \sum_{\mu} \left( a^{\dg}_{\mu} F_{\mu}
       + \rm{H.c.}\right),
\end{eqnarray}
%%\end{subequations}
then, by considering this tunneling Hamiltonian as a perturbation,
the second-order Born expansion results in an {\it unconditional}
master equation for the reduced density matrix
of the same form of \Eq{ME-1}.

To construct a {\it conditional} (i.e. particle-number-resolved)
master equation, one should keep track of the number
of electrons that arrive at the collector.
%%%%
Let us classify the Hilbert space of the electrodes as follows.
First, we define the subspace in the absence of electrons
 at the collector as
``$B^{(0)}$", which is spanned by the product
of all many-particle states of the two isolated reservoirs, formally denoted
as $B^{(0)}\equiv\mb{span}\{|\Psi_L\ra\otimes |\Psi_R\ra \}$.
%%%%
Then, we introduce the Hilbert subspace ``$B^{(n)}$" ( $n=1,2,\cdots$),
corresponding to $n$-electrons at the collector.
The entire Hilbert space of the two electrodes is $B=\oplus_n B^{(n)}$.

With this type of classification for the reservoir states,
the average over states in the entire Hilbert space ``$B$"
is replaced with states in the subspace ``$B^{(n)}$",
leading to a {\it conditional} master equation \cite{Li05b}
\bea\label{ME-2a}
\dot{\rho}^{(n)}(t) &=& -i{\cal L}\rho^{(n)}(t) - \int^{t}_{0}d\tau
      \mb{Tr}_{B^{(n)}} [{\cal L}'(t){\cal G}(t,\tau)   \nl
      & &  \times {\cal L}'(\tau) {\cal G}^{\dg}(t,\tau) \rho_T(t)] .
\eea
Here, $\rho^{(n)}(t)=\mb{Tr}_{B^{(n)}}[\rho_T(t)]$,
where $\rho_T(t)$ is the state of the whole system.
$\rho^{(n)}(t)$ is the reduced density matrix of the {\it central} system
conditioned on the number of electrons arriving at the collector
until time $t$.

As for the qubit-QPC problem, two important considerations
are made, as follows:
(i)
Instead of the conventional Born approximation for the entire density matrix
$\rho_T(t)\simeq\rho(t)\otimes\rho_B$,
we propose the ansatz $\rho_T(t)\simeq\sum_n\rho^{(n)}(t)\otimes\rho_B^{(n)}$,
where $\rho_B^{(n)}$ is the density operator of the electron reservoirs
associated with $n$-electrons arrived at the collector.
(ii)
Due to the closed nature of the transport circuit, the extra
electrons arriving at the collector will flow back into the emitter
(left reservoir) via the external circuit. In addition, the rapid
relaxation processes in the reservoirs will quickly bring the
reservoirs to the local thermal equilibrium state determined by
the chemical potentials.

Then, under the Markovian approximation, from \Eq{ME-2a}, we obtain \cite{Li05b}
\bea\label{ME-3a}
\dot{\rho}^{(n)}
   &=&  -i {\cal L}\rho^{(n)} - \frac{1}{2} \sum_{\mu}
   \left\{ [a_{\mu}^{\dg}A_{\mu}^{(-)}\rho^{(n)}
   +\rho^{(n)}A_{\mu}^{(+)}a_{\mu}^{\dg}
\right. \nl & &         - A_{L\mu}^{(-)}\rho^{(n)}a_{\mu}^{\dg}
            - a_{\mu}^{\dg}\rho^{(n)}A_{L\mu}^{(+)}        \nl
& & \left. - A_{R\mu}^{(-)}\rho^{(n-1)}a_{\mu}^{\dg}
   - a_{\mu}^{\dg}\rho^{(n+1)}A_{R\mu}^{(+)}]+{\rm H.c.} \right\} .
\eea
Here, $A_{\alpha\mu}^{(\pm)}=\sum_{\nu}
C_{\alpha\mu\nu}^{(\pm)}(\pm {\cal L})a_{\nu}$
and $A_{\mu}^{(\pm)}=\sum_{\alpha=L,R}A_{\alpha\mu}^{(\pm)}$.
The spectral functions $C_{\alpha\mu\nu}^{(\pm)}(\pm {\cal L})$
are defined in terms of the Fourier transform of the reservoir
correlation functions, i.e.,
$ C_{\alpha\mu\nu}^{(\pm)}(\pm {\cal L})=\int^{\infty}_{-\infty} dt
  C_{\alpha\mu\nu}^{(\pm)}(t) e^{\pm i{\cal L}t}$,
where
$\la f^{\dg}_{\alpha\nu}(\tau) f_{\alpha\mu}(t)\ra
= C_{\alpha\mu\nu}^{(+)}(t-\tau)$
and $\la f_{\alpha\mu}(t)f^{\dg}_{\alpha\nu}(\tau)\ra
= C_{\alpha\mu\nu}^{(-)}(t-\tau)$.

The ``$n$"-dependence of \Eq{ME-3a} is somehow similar to the usual
rate equation, despite its operator feature.
Each term in \Eq{ME-3a} can be interpreted similarly using
the conventional {\it c-number} rate equation.
%%%%
Unlike in the Bloch equation derived by Gurvitz
{\it et al} \cite{Gur96b}, in \Eq{ME-3a}, $\rho^{(n)}$
is also coupled to $\rho^{(n+1)}$, which is not present in
Ref.\ \onlinecite{Gur96b}.
This difference originates from the fact
that \Eq{ME-3a} is valid for non-zero temperatures.

\subsection{Current and Noise Spectrum}

\subsubsection{Qubit measurement by QPC }

{\flushleft With the knowledge of }
$\rho^{(n)}(t)$, one can carry out
the various readout characteristics of the measurement.
In particular, for a qubit with $\Omega\neq 0$ and
$\Omega^{-1}$ comparable to
or smaller than the measurement time \cite{Sch98},
the qubit oscillation cannot be read out
using the conventional single shot measurement.
In this regime, continuous weak measurement is
an alternative scheme to record the qubit oscillations,
e.g., in the power spectrum of the output current.

First, for the ensemble-average current,
simple expression is related to the unconditional
density matrix $\rho(t)=\sum_n\rho^{(n)}(t)$.
%%%%%%
The derivation is from the fact that the current
is associated with the probability distribution
$P(n,t)\equiv\rm{Tr}[\rho^{(n)}(t)]$, via
$I(t)=ed \bar{N}(t) /dt$, where $\bar{N}(t)=\sum_n nP(n,t)$.
By considering the Summation over ``$n$" and making use of the
cyclic property under trace, we obtain \cite{Li05}
\bea\label{it}
I(t) =  \sum_n n
\mb{Tr}[\dot{\rho}^{(n)}(t)]
     = \frac{1}{2}\mb{Tr}[\bar{Q}\rho(t) Q + \mb{H.c.} ] ,
\eea
where $\bar{Q} = \ti{Q}^{(-)}-\ti{Q}^{(+)}$.

Second, the power spectrum of the output current can be
conveniently calculated using the MacDonald's formula \cite{Gur03,Li05}
\bea \label{Mac-1}
S(\omega) = 2\omega \int^{\infty}_{0} dt \sin \omega t \frac{d}{dt}
            \left[\la n^2(t)\ra - (\bar{I} t)^2\right] ,
\eea
where $\bar{I}$ is the average current over time
and $\la n^2(t)\ra = \sum_n n^2 P(n,t)$.
It can be proved that \cite{Li05}
\bea \label{Mac-2}
\frac{d}{dt} \la n^2(t)\ra = \mb{Tr}\left[ \bar{Q} \hat{N}(t)Q
             + \frac{1}{2}\ti{Q}\rho(t)Q + \mb{H.c.}\right] ,
\eea
where $\hat{N}(t)\equiv \sum_n n \rho^{(n)}(t)$,
which can be calculated via its equation of motion \cite{Li05}
\bea \label{Mac-3}
\frac{d\hat{N}}{dt} = -i {\cal L}\hat{N}
   -\frac{1}{2}\left[Q,\ti{Q}\hat{N}-\hat{N}\ti{Q}^{\dg} \right]
   + \frac{1}{2}(\bar{Q}\rho Q +\mb{H.c.}) \,.  \nl
\eea
Combining the above three equations, the noise spectrum
$S(\omega)$ can be easily obtained
via Laplace transform
in terms of simple algebraic manipulations.

\subsubsection{Quantum transport}

{\flushleft For quantum transport}, based on the $n$-ME
(\ref{ME-3a}), a method similar to that outlined above leads to \cite{Li05b}
\bea\label{I-t}
&& I(t) = e \sum_n n \rm{Tr} \left[ \dot{\rho}^{(n)}(t)\right] \nl
     && = \frac{e}{2} \sum_{\mu} \rm{Tr} \left[ \left(
         a^{\dg}_{\mu}A^{(-)}_{R\mu}-A^{(+)}_{R\mu}a^{\dg}_{\mu}
         \right)\rho(t)+\rm{H.c.} \right] .
\eea
Here, the unconditional density matrix $\rho=\sum_n \rho^{(n)}$
satisfies the usual master equation
(which can be obtained by summing up \Eq{ME-3a} over ``$n$")
\bea\label{rho-t}
 \dot{\rho} = -i {\cal L}\rho - \frac{1}{2}\sum_{\mu}
        \left\{ [a_{\mu}^{\dg},A_{\mu}^{(-)}\rho
        -\rho A_{\mu}^{(+)}]+ \rm{H.c.} \right\}  .
\eea
\Eqs{I-t} and (\ref{rho-t}) can serve as a convenient
starting point to compute the transport current.
In practice, one can first diagonalize the central
system Hamiltonian, then perform the Liouvillian
operations in the eigen-state representation.

For transport, the current noise spectrum
can provide additional {\it dynamic}
information beyond the current itself.  
We know that, for time-dependent transport, the currents across
the left and right junctions (between the central system
and the two leads) are not necessarily equal to each other.
This requires a definition for the power spectrum
using the ``average" current $I(t)=\alpha I_L(t)+\beta I_R(t)$,
where $\alpha$ and $\beta$ are two coefficients determined by the
junction capacitances \cite{Li053}, which satisfy $\alpha+\beta=1$.
This leads to the noise spectrum of the current consisting of
three parts \cite{Li053}:
%\begin{eqnarray}\label{Sc}
$ S(\omega)=\alpha S_L(\omega)+\beta S_R(\omega)
-\alpha\beta\omega^2S_N(\omega)$,
%\end{eqnarray}
where $S_{L/R}(\omega)$ is the noise spectrum of the left (right)
junction current $I_{L/R}(t)$
and $S_N(\omega)$ is the fluctuation spectrum of the electron number
$N(t)$ on the central device.

For $S_\lambda(\omg)$ ($\lambda=L,R$), we use the MacDonald's formula
 \be\label{MacD}
 S_\lambda(\omega)=2\,\omg\!\int_{0}^{\infty}\!\!dt\sin(\omg t)
 \frac{d}{dt}
 [\la n_\lambda^{2}(t)\ra-(\bar{I}t)^2] \,,
 \ee
 where $\bar{I}$ is the stationary current and
 $\la n_\lambda^{2}(t)\ra=\sum_{n_\lambda}n_\lambda^{2}
 [\mb{Tr}\rho^{(n_\lambda)}(t)]
 =\sum_{n_\lambda}n_\lambda^{2}P(n_{\lambda},t)$.
Using \Eq{ME-3a}, we obtain \cite{Li053}
%%%%%
 \be\label{dn2dt}
 \frac{d}{dt}\la n_\lambda^{2}(t)\ra
 =\mb{Tr}\big[2\mathcal{T}^{(-)}_\lambda N^\lambda(t)
 +\mathcal{T}^{(+)}_\lambda\rho^{\rm st}\big],
\ee
where $N^\lambda(t)\equiv\sum_{n_\lambda}n_\lambda\rho^{(n_\lambda)}(t)$
 denotes the ``number'' matrix and
 $\rho^{\rm st}$ is the stationary state.
Here, we also introduce
\be
 \mathcal{T}^{(\pm)}_\lambda(\cdots)
 =\frac{1}{2}\sum_\mu\big[A_{\lambda\mu}^{(-)}
  (\cdots)a_\mu^\dag \pm
  a_\mu^\dag(\cdots)A_{\lambda\mu}^{(+)}\big]\!+\!\mb{H.c.}
 \label{calTpm}.
 \ee
The final expression for $S_{\rm L/R}(\omg)$ is \cite{Li053}
 \bea\label{Salpha}
 S_\lambda(\omg)
 \!&=&\!4\,\omg\,{\rm Im}\big\{ {\rm Tr} \big[
 {\cal T}^{(-)}_\lambda\wit{N}^\lambda(\omg)\big]\}
 \nl
 &&+\,2\,{\rm Tr}\big[{\cal T}^{(+)}_\lambda\rho^{\rm st}\big]
 -8\,\pi\,\bar{I}^2\,\delta(\omg),
\eea%
where $\wit{N}^\lambda(\omg)=\int_0^\infty dtN^\lambda(t) e^{i\omg t}$.
The last term originates from the
second term in the MacDonald's formula in \Eq{MacD}.
$\wit{N}^\lambda(\omg)$ can be easily obtained by solving
the following equation of motion in the frequency domain \cite{Li053}:
\be\label{dNdt}
 \frac{d}{dt}N^\lambda(t)=-i{\cal L}N^\lambda(t)
  -{\cal R}N^\lambda(t)+{\cal T}^{(-)}_\lambda\rho^{\rm st},
\ee
which gives
\bea\label{Nomega}
 \!\!\!\!-i\omg \wit{N}^\lambda(\omg)
 \!\!\!&=&\!\!
 -i{\cal L}\wit{N}^\lambda(\omg)\!-\!{\cal R}\wit{N}^\lambda(\omg)
 \!+\!{\cal T}^{(-)}_\lambda\wit{\rho}(\omg),
\eea%
where $\wit{\rho}(\omg)=i\rho^{\rm st}/\omg$.

For the charge fluctuations on the central system,
the symmetrized noise spectrum can be expressed using \cite{Li053}
 \be
 S_N(\omg)=\omg^2\! \int_{-\infty}^{\infty}\!d\tau
 \la \hat{N}(\tau)\hat{N}+\hat{N}\hat{N}(\tau) \ra e^{i\omg\tau},
 \ee
where $\la \hat{N}(\tau) \hat{N} \ra=\mb{Tr}\mb{Tr}_{\rm B}
[U^{\dag}(\tau)\hat{N}U(\tau)\hat{N}\rho^{\rm st}\rho_{\rm B} ]$, where
$U(\tau)=e^{-iH\tau}$ and $\hat{N}$
is the electron-number operator of the central system.
%%%%
Using the cyclic property under trace, we have
$\la \hat{N}(\tau)\hat{N} \ra=
\mb{Tr}[\hat{N}\sgm(\tau)]$, where $\sgm(\tau)= \mb{Tr}_{\rm B}
[U(\tau)\hat{N}\rho^{\rm st}\rho_{\rm B} U^{\dag}(\tau)]$.
Desirably, $\sgm(\tau)$ satisfies the equation of the usual
reduced density matrix.
Its Fourier transform $\wit{\sgm}(\omg)$
can be easily solved using \cite{Li053}
 \bea\label{rhoomega}
 i(\omg-{\cal L})\wit{\sgm}(\omg)
 ={\cal R}\wit{\sgm}(\omg)
 -\hat{N}\rho^{\rm st}.
 \eea
%%%%
Then, we have \cite{Li053}
\bea\label{SNw}
S_N(\omg)=2\omg^2\mb{Re}\{ {\rm Tr}
\{\hat{N}[\wit{\sgm}(\omg)+\wit{\sgm}(-\omg)]\} .
\eea

\subsection{Counting Statistics and Large-Deviation Analysis}

{\flushleft In order to get}
information in addition to the average current
and current fluctuation spectrum,
with the knowledge of $\rho^{(n)}(t)$ (and thus $P(n,t)$),
one can perform FCS \cite{Lev9396}
and large-deviation (LD) analysis \cite{LD9809,SM8792,Gar10}.
For FCS analysis, the current CGF
can be constructed using \cite{FCS07}
\begin{equation}\label{CGF}
e^{-\tilde{\cal F}(\chi,t)}=\sum_n P(n,t)e^{in\chi} ,
\end{equation}
where $\chi$ is the so-called counting field.
Based on the CGF, $\tilde{\cal F}(\chi,t)$,
the $k_{\rm th}$ cumulant can be readily carried out via
$C_k=-(-i\partial_{\chi})^k \tilde{\cal F}(\chi,t)|_{\chi=0}$.
As a result, one can easily confirm that
the first two cumulants, $C_1=\bar{n}$ and $C_2=\overline{n^2}-\bar{n}^2$,
give rise to the mean value and the variance of the transmitted electrons,
while the third one (skewness), $C_3=\overline{(n-\bar{n})^3}$,
characterizes the asymmetry of the distribution.
Here, $\overline{(\cdots)}=\sum_n (\cdots)P(n,t)$.
Moreover, one can relate the cumulants to measurable quantities,
e.g., the average current by $I =eC_{1}/t$,
and the zero-frequency shot noise by $S =2e^{2}C_{2}/t$.
In addition, the important Fano factor is simply given by $F=C_{2}/C_{1}$,
which characterizes the extent of current fluctuations:
$F>1$ indicates a super-Poisson fluctuating behavior,
while $F<1$ indicates a sub-Poisson process.

For the LD analysis,
instead of using the discrete Fourier transform in \Eq{CGF},
one may consider $e^{i\chi n}\Rightarrow e^{-xn}$.
That is, introduce the {\it dual}-function of $P(n,t)$ \cite{LD11}:
\bea\label{LD-1}
P(x,t)=\sum_n e^{-x n} P(n,t)=e^{-{\cal F}(x,t)}.
\eea
The real nature of the transform factor $e^{-x n}$,
in contrast to the complex one, $e^{i\chi n}$,
makes the resultant $P(x,t)$ somewhat resemble the
partition function in statistical mechanics.
Using analogous terms in statistical mechanics,
in \Eq{LD-1}, the trajectories are categorized
by a dynamical order parameter ``$n$" or its conjugate field ``$x$".
This is realized by an exponential weight similar to the Boltzmann factor,
with the dynamical order parameter representing the energy or magnetization
and the conjugate field representing the temperature or magnetic field.

$P(x,t)$ is called {\it LD function} in LD analysis. 
In statistical mechanics, the partition function measures the number of
microscopic configurations accessible to the system under given conditions.
For the mesoscopic transport under consideration, if we are interested
in the dynamical aspects of the transport electrons, the above insight
can be used for LD analysis in the {\it time domain}.
That is, the LD function is a measure of the number of trajectories accessible
to the ``counter", which favorably characterizes the trajectory space
from multiple angles according to the effect of the conjugate field.
In particular, it allows one to inspect the {\it rare fluctuations}
or {\it extreme events} by tuning the conjugate field ``$x$".

We emphasize that, if one performs the conventional FCS analysis only,
using either a complex transform factor $e^{i\chi n}$ or a real one $e^{-x n}$
would make no difference since the limit
$ \chi (x) \rightarrow 0$ will be considered at the end.
However, for the LD study, we must use the real factor $e^{-x n}$,
which plays a role in categorizing (selecting) the trajectories.
This type of selection would enables us to perform statistical
analysis for the fluctuations of {\it sub-ensembles} of trajectories.
For instance, $x>0$ implies mainly selecting the {\it inactive} trajectories
(with small $n$),
while $x<0$ prefers the {\it active} trajectories (with large $n$).
In particular, by varying $x$, the $x$-dependent statistics can
reveal interesting dynamical behaviors in the {\it time-domain}.
In other words, based on the distribution function $P(n,t)$,
which contains complete information of all the trajectories,
the LD approach, beyond the conventional FCS analysis, captures more
information from $P(n,t)$ via the $x$-dependent cumulants.

From the technical point of view,
similar to transforming $P(n,t)$ to $P(x,t)$,
we introduce $\rho(x,t)=\sum_n e^{-xn}\rho^{(n)}(t)$.
Then, from \Eq{nME},  we formally have \cite{LD11}
\bea\label{rho-xt}
\dot{\rho}(x,t) = \left[ -i{\cal L} - {\cal R}_0
  - e^{x}{\cal R}_{1} - e^{-x}{\cal R}_{-1} \right] \rho(x,t) .
\eea
This equation allows us to carry out the LD function $P(x,t)$
via $P(x,t)={\rm Tr}[\rho(x,t)]$,
where the trace is over the central system states.
Accordingly, we obtain the generating function
${\cal F}(x,t)=-\ln P(x,t)$ for {\it arbitrary} counting time $t$.
Further, we can prove \cite{LD11}
\begin{subequations}\label{F-k}
\begin{align}
{\cal F}_1(x,t)&\equiv \partial_x {\cal F}(x,t)
=\frac{1}{P(x,t)}\sum_n n e^{-xn} P(n,t)
  \equiv \la n \ra_x ,   \\
{\cal F}_2(x,t)&\equiv \partial^2_x {\cal F}(x,t)
=-\la (n-\bar{n}_x )^2\ra_x,
\end{align}
and more generally,
\begin{align}
{\cal F}_k(x,t)\equiv \partial^k_x {\cal F}(x,t)
=(-)^{(k+1)}\la (n-\bar{n}_x )^k\ra_x.
\end{align}
\end{subequations}
Here, for brevity, we used the notation $\bar{n}_x$ for $\la n \ra_x$.
Using these cumulants, we can define a finite-counting-time
average current $I(x,t)=e{\cal F}_1(x,t)/t$ and the
shot noise $S(x,t)=2e^2 {\cal F}_2(x,t)/t$.
In addition, the generalized Fano factor, $F(x,t)={\cal F}_2(x,t)/{\cal F}_1(x,t)$,
will be of interest to characterize the fluctuation properties.

We notice that, in order to obtain ${\cal F}_k(x,t)$,
we only need to determine the various $k$-th order derivatives of $P(x,t)$,
$P_k(x,t)=\partial^k_x P(x,t)$.
This is an efficient method to
compute the $x$-dependent cumulants for finite counting time.
That is, by performing the derivatives $\partial^k_x$ on \Eq{rho-xt}
and defining $\rho_k(x,t)=\partial^k_x \rho(x,t)$,
we obtain a set of coupled equations for $\rho_k(x,t)$,
whose solution then gives $P_k(x,t)={\rm Tr}[\rho_k(x,t)]$.

In long counting time limit, it can be proved that
${\cal F}(x,t)\simeq t\lambda(x)$.
Desirably, the asymptotic form, $P(x,t)\simeq e^{-t\lambda(x)}$,
allows one to identify the LD function $\lambda(x)$
for the {\it smallest} eigenvalue
of the counting matrix, i.e., the r.h.s of \Eq{rho-xt}.

% =======================================================
\section{Application to Quantum Measurement}

{\flushleft In this section, }
we illustrate the application of the $n$-ME
to two examples of quantum measurement,
i.e., for a charge qubit measured respectively
by QPC and SET detectors.

\subsection{Qubit Measured by QPC}

{\flushleft Let us specify}
the charge qubit as a pair of coupled quantum dots,
described by the Hamiltonian
$ H_{\rm qu} =  \epsilon_a |a\ra\la a| + \epsilon_b |b\ra\la b|
+ \Omega(|b\ra\la a|+|a\ra\la b|) $.
We then introduce $\epsilon=(\epsilon_a-\epsilon_b)/2$
and set $(\epsilon_a+\epsilon_b)/2$ as the reference energy.
%%%%
The qubit eigen-energies are obtained as
$E_1=\sqrt{\epsilon^2+\Omega^2}\equiv \Delta/2$
and $E_0=-\sqrt{\epsilon^2+\Omega^2}=-\Delta/2$.
Correspondingly, the eigenstates are
$|1\ra=\cos\frac{\theta}{2}|a\ra+\sin\frac{\theta}{2}|b\ra$
for the excited state
and $|0\ra=\sin\frac{\theta}{2}|a\ra-\cos\frac{\theta}{2}|b\ra$
for the ground state, where $\theta$ is introduced using
$\cos\theta=2\epsilon/\Delta$ and $\sin\theta=2\Omega/\Delta$.
%%%%%%%
The coupling between the qubit and detector is characterized by
$H'=QF$, where $Q = {\cal T}+\chi |a\ra\la a|$
and $F=\sum_{k,q} (c^{\dg}_kd_q + \mb{H.c.})$.

By applying \Eq{it},
we obtain the stationary current for a
symmetric qubit ($\epsilon=0$) as \cite{Li05}
\bea\label{Iss}
I_{s}=g_0V+g_1V\left[1-2\frac{G^{(-)}}{V}
       + \frac{\Delta}{V}\frac{G^{(-)}}{G^{(+)}} \right] .
\eea
Here, $g_0=\eta ({\cal T}+\chi/2)^2$, $g_1=\eta (\chi/2)^2$, and
$ G^{(\pm)} = \frac{1}{2}\left[F^{(+)}(\Delta,V)
                        \pm F^{(-)}(\Delta,V) \right] $,
with
$F^{(\pm)}(\Delta,V)\equiv (\Delta\pm V)\coth(\frac{\Delta\pm V}{2T})$.
%%%%%%%
At zero temperature, \Eq{Iss} can be further simplified.
Compared with previous results \cite{Shn02,Sta03},
we find that under low voltage ($V<\Delta$), \Eq{Iss} is reduced
to the same result given by Shnirman {\it et al.} \cite{Shn02},
but under $V>\Delta$ it differs from the results in
Refs.\ \onlinecite{Shn02} and \onlinecite{Sta03}.

The output power spectrum $S(\omega)$
can be calculated using the MacDonald's formula
via Eqs.\ (\ref{Mac-1})-(\ref{Mac-3}).
For a symmetric qubit and denoting
$S(\omega)=S_0+S_1(\omega)+S_2(\omega)$, we obtain \cite{Li05}
\begin{subequations}\label{SW}
\bea
S_0 &=& 2I_0 \coth\frac{V}{2T} + \frac{\chi^2 \eta}{2} \nl
          & &    \times \left[ G^{(+)}-\frac{\Delta^2}{G^{(+)}}
                 -V\coth\frac{V}{2T} \right] ,\\
S_1(\omega) &=& \left[1-\frac{\Delta}{2V}\frac{G^{(-)}}{G^{(+)}} \right]
       \frac{I^2_d\Gamma_d\Delta^2}{(\omega^2-\Delta^2)^2
       +\Gamma_d^2\omega^2} ,\\
S_2(\omega) &=& \chi^2 \eta \left[\Gamma_d D_z
             +\gamma \bar{I} \right]
                \frac{G^{(-)}}{\omega^2+\Gamma_d^2}  .
\eea
\end{subequations}
Here, three currents are defined as $I_0=(I_a+I_b)/2$, $I_d=I_a-I_b$,
and $\bar{I}= I_0-\frac{1}{4} \eta\chi^2\Delta G^{(-)}/G^{(+)}$,
with $I_a=\eta ({\cal T}+\chi)^2V$ and $I_b=\eta {\cal T}^2V$ being
the detector currents corresponding to
qubit states $|a\ra$ and $|b\ra$, respectively.
%%%%%%
The other quantities in \Eq{SW} are
$\Gamma_d=\frac{\eta\chi^2}{2}G^{(+)}$,
$\gamma=\frac{\eta\chi^2}{2}\Delta$
and $D_z= -\Delta\sqrt{I_aI_b}/G^{(+)}-\eta\chi^2 G^{(-)}/4$.
%%%%%%
The three noise spectrum components are, respectively,
(i) the zero-frequency noise $S_0$,
(ii) the Lorentzian spectral function $S_1(\omega)$
with a peak around the qubit Rabi frequency $\omega=\Delta$,
and (iii) $S_2(\omega)$, originating from the qubit-relaxation-induced inelastic tunnelling effect in the detector.
In addition to $S_2(\omega)$, the qubit relaxation effect is also manifested
in $S_0$ and $S_1(\omega)$,
i.e., giving rise to the second term of $S_0$
and reducing the pre-factor in $S_1(\omega)$ from unity.
%%%%
If the qubit-relaxation-induced inelastic effect is neglected
or at the limit of the high bias voltage $V\gg\Delta$,
\Eq{SW} returns to the result obtained in previous work
\cite{Kor01a,Goa01a}.

\begin{figure}[h]
%\begin{center}
\centerline{\includegraphics [scale=0.3,angle=0] {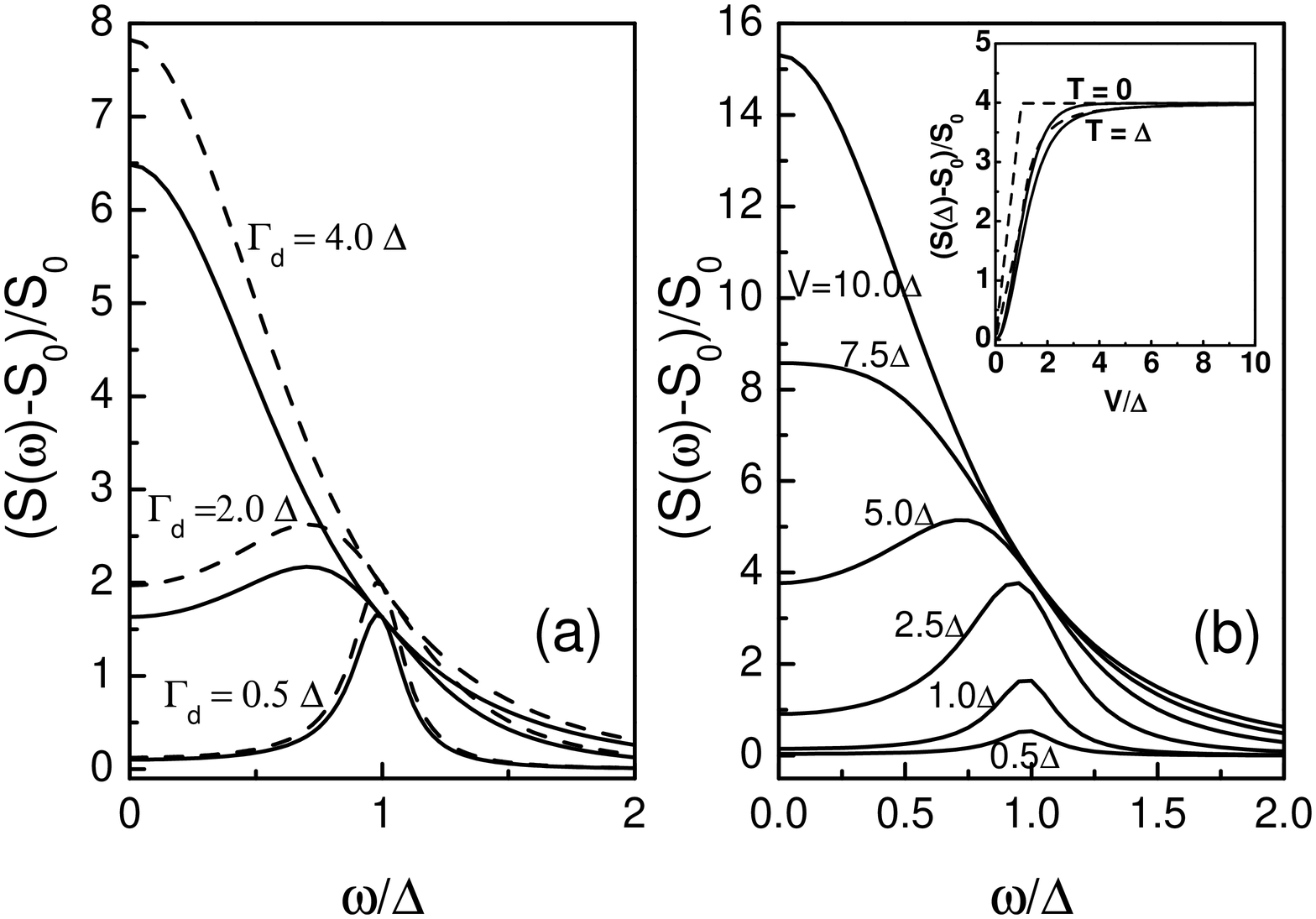}}
%\end{center}
\caption{(a) Noise spectrum in the presence (solid curves) and
absence (dashed curves) of qubit relaxation.
(b) Voltage effect on the noise spectrum, particularly on the
peak-to-pedestal ratio (inset, where the solid and dashed curves
correspond to the presence and absence of qubit relaxation).
%%%%%%%%
The results in (a) and (b) are obtained, respectively, by altering
$\chi$ (for a fixed voltage $V=2\Delta$ ) and the voltage $V$
(for a fixed $\chi=0.1\Delta$).
Other parameters: $g_L=g_R=2.5/\Delta$, and $T=\Delta$. }
\end{figure}

\begin{figure}[h]
%\begin{center}
\centerline{\includegraphics [scale=0.5,angle=0] {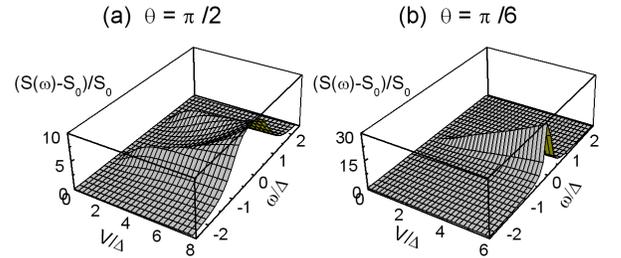}}
%\end{center}
\caption{ 3D-plot of the noise spectra for (a) the symmetric
and (b) the asymmetric qubit.
Parameters: $g_L=g_R=2.5/\Delta$, $\chi=0.1\Delta$,
and $T=\Delta$. }
\end{figure}

The measurement-induced relaxation effects of the qubit are shown
in Fig.\ 3. The major effect of the qubit relaxation shown in Fig.\ 3(a)
is the lowering of the entire noise spectrum in qualitative consistence
with the findings of Gurvitz {\it et al} \cite{Gur03}, where an
{\it external thermal bath} is introduced to cause qubit relaxation.
However, the {\it spontaneous} relaxation discussed here
does not diminish the telegraph noise peak near zero frequency
in the incoherent case,
which implies the presence of the Zeno effect,
in contrast to the major conclusion of Ref.\ \onlinecite{Gur03}.
In addition, the transition behavior from the coherent to the incoherent regime
is different.
%%%%%%%%
Figure 3(b) shows the voltage effect where
the coherent peak around $\omega=\Delta$ reduces as
the measurement voltage decreases.
Interestingly, this effect alters the fundamental bound of 4
for the signal-to-noise ratio, 
$[S(\Delta)-S_0]/S_0$,
which was determined by Korotkov {\it et al.} at the high voltage limit
(see the inset) \cite{Kor01a}.

The voltage effect is further shown in Fig.\ 4 by the
3D plot of the scaled spectra for different qubit symmetries.
%%%%%%
In contrast to the present result,
we notice that in Ref.\ \onlinecite{Sta03}, no spectral structure was found,
i.e., $S(\omega)-S(\infty)=0$ in the range of $V<10\Delta$
for the symmetric qubit ($\theta=\pi/2$).
However, Shnirman {\it et al} showed the existence of the coherent peaks
at $\omega=\pm\Delta$ for voltage greater than $\Delta$
\cite{Shn02}.
%%%%%%%
For an asymmetric qubit, as shown in Fig.\ 4(b), the coherent peaks at
$\omega=\pm\Delta$ are destroyed and a peak around $\omega=0$ is formed.
This transition originates from the breakdown of the resonant condition,
which replaces the Rabi oscillation of the qubit with incoherent jumping.

\subsection{Qubit Measured by SET}

{\flushleft As second example of quantum measurement},
we consider a charge qubit measured
by an SET \cite{Dev00,Sch98,Sch01},
as schematically shown in Fig.\ 5.
The SET is a sensitive charge-state detector, which is suitable for
fast qubit read-out in solid-state quantum computation.
For single-shot measurement, i.e., where
the qubit state is unambiguously determined in one run,
an important figure of merit is the detector's efficiency,
defined as the ratio of information gained time
and the measurement-induced dephasing time \cite{Sch98,Sch01}.
In the weakly responding regime, it was found that the SET has
a rather poor quantum efficiency \cite{Sch01,Kor01,Moz04}.
However, later study showed that, for a strong-response SET,
the quantum limit of an ideal detector can be reached,
resulting in an almost pure conditioned state \cite{Wis06}.

\begin{figure}[h]
\begin{center}
\includegraphics[width=7cm]{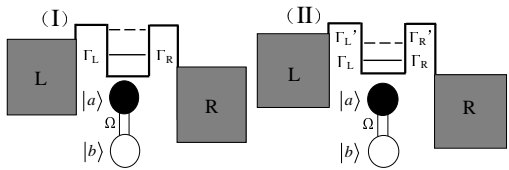}
\caption{\label{fig1} Schematics for a solid-state qubit
measurement by SET. Model (I): the SET level is within the bias
window for qubit state $|b\ra$, but outside for state
$|a\ra$. Model (II): the SET level is between the Fermi levels
for either $|b\ra$ or $|a\ra$, but with different couplings
to the leads, i.e., $\Gamma_{L/R}$ for $|b\ra$, and
$\Gamma'_{L/R}$ for $|a\ra$. }
\end{center}
\end{figure}

As mentioned earlier for the QPC detector,
a more implementable approach is continuous weak measurement rather than single-shot measurement.
This type of measurement allows one to determine the ensemble average of
detector and qubit states,
and the qubit coherent oscillation is read out from the
spectral density of the detector.
In continuous weak measurement, an interesting generic result is
the so-called Korotkov-Averin (K-A) bound, i.e., the signal-to-noise
ratio (SNR) bounded by a fundamental limit of ``4" \cite{K-A01},
which can be broken only in cases
such as when performing quantum nondemolition (QND) measurement \cite{Ave02},
adding quantum feedback control \cite{Wang07},
or using two detectors \cite{But06}.
We consider continuous weak measurement of qubits
using strongly responding SETs \cite{Wis06,Gur05}
and show that, for both models in Refs.\ \onlinecite{Wis06}
and \onlinecite{Gur05}, the SNR
can violate the \emph{universal} Korotkov-Averin bound \cite{AK}.

The entire method of the qubit-SET measurement
is described by the following Hamiltonian \cite{Wis06,Gur05,AK}
\begin{subequations}
\begin{align}
H &= H_0+H^\prime\\
H_0 &= H_S+\sum_{\lambda=L,R}\epsilon_{\lambda k}d_{\lambda k}^\dag
d_{\lambda k}   \label{Eq1}\\
%H_S &= \sum_{i=a,b,c}E_ia_i^\dag a_i+\Omega(a_a^\dag a_b+a_b^\dag a_a)
% +Un_an_c\\
H_S &= \sum_{j=a,b}E_j|j\ra\la j|+\Omega(|a\ra\la b|+|b\ra\la a|)
 + E_c a_c^\dag a_c +Un_an_c\\
H^\prime &=\sum_{\lambda=L,R;k}(\Omega_{\lambda k}a_c^\dag d_{\lambda
k}+{\rm H.c.}) \equiv ~ a_c^\dag (f_{cL}+f_{cR})+ {\rm H.c.}
\end{align}
\end{subequations}
For simplicity, we assumed spinless electrons.
The system Hamiltonian, $H_S$, contains a qubit, SET central dot,
and their Coulomb interaction (the $U$-term).
For the qubit, we assumed that each dot has only one bound state,
i.e., the logic states $|a\ra$ and $|b\ra$ with energies $E_a$
and $E_b$ and  a coupling amplitude $\Omega$.
$n_a$ is the number operator of qubit state $|a\ra$, which
is 1 for $|a\ra$ being occupied and 0 otherwise.
For the SET, $a_c^\dag (a_c)$ and $d_{\alpha k}^\dag (d_{\alpha k})$ are
the electron creation (annihilation) operators of the central dot and
reservoirs, respectively. $n_c\equiv a_c^\dag a_c$ is introduced as the number operator
of the SET dot. Similar to previous work,
we assumed that the SET works in the strong Coulomb-blockade regime,
with only a single level $E_c$ involved in the measurement process.
Finally, $H'$ describes the tunnel coupling of the SET dot to the leads,
with amplitude $\Omega_{\lambda k}$.

We consider the two models schematically shown in Fig.\ 5.
In model (I), which was studied in Ref.\ \onlinecite{Gur05},
the SET dot level is within the bias voltage
if the qubit is in state $|b\ra$, but is above the Fermi level
when the qubit state is switched to $|a\ra$.
For state $|b\ra$, a nonzero current $I_b$ flows through
the SET; however, for state $|a\ra$, the SET current $I_a$ is zero.
Then, the qubit state can be discriminated from these different currents.
In this model, the signal current $\Delta I\equiv |I_b-I_a|$ is
twice the average current $\bar{I}\equiv (I_b+I_a)/2$.
Hence, it is not a weak response detector.
In model (II), which
illustrates the crossover from weak to strong responses,
the SET dot level is always between the
Fermi levels of the two leads for qubits either in state
$|b\ra$ or state $|a\ra$, but with different coupling strengths
to the leads, i.e., $\Gamma_{L(R)}$ and $\Gamma'_{L(R)}$.
We further
parameterize the tunnel couplings as
$\Gamma_L(\Gamma_L^\prime)=(1\pm\xi)\bar{\Gamma}_L$,
$\Gamma_R(\Gamma_R^\prime)=(1\pm\zeta)\bar{\Gamma}_R$,
and $\gamma=\bar{\Gamma}_R/\bar{\Gamma}_L$.
Here, $\bar{\Gamma}_{L(R)}=(\Gamma_{L(R)}+\Gamma'_{L(R)})/2$
denotes the average couplings,
while $\xi$ and $\zeta$ characterize the response strength of the
detector to qubits. In this context,
we would like to mention that, usually, the analysis is
restricted in the weak-response regime by assuming that
$\xi\ll 1$ and $\zeta\ll 1$, except in Ref.\ \onlinecite{Wis06},
where the quantum efficiency was investigated
in the strong response regime using this model.

For the both models in Fig.\ 5, the states involved
are  $|1\rangle=|0a\rangle$,
$|2\rangle=|0b\rangle$, $|3\rangle=|1a\rangle$,
and $|4\rangle=|1b\rangle$.
In this notation $|0(1)a(b)\rangle$ means that the SET dot
is empty (occupied) and the qubit is in state $|a(b)\rangle$.
Applying \Eq{ME-3a} to model (I) yields \cite{AK}
\begin{subequations}
\begin{align}
\dot{\rho}^{(n_R)}_{11}=&i\Omega[\rho^{(n_R)}_{12}-\rho^{(n_R)}_{21}]
+\Gamma_L\rho^{(n_R)}_{33}+\Gamma_R\rho^{(n_R-1)}_{33}\\
%%%%%%%%%%
\dot{\rho}^{(n_R)}_{22}=&i\Omega[\rho^{(n_R)}_{21}-\rho^{(n_R)}_{12}]
-\Gamma_L\rho^{(n_R)}_{22}+\Gamma_R \rho^{(n_R-1)}_{44}\\
%%%%%%%%%%%%%
\dot{\rho}^{(n_R)}_{12}=&-i\epsilon\rho^{(n_R)}_{12}+i\Omega[\rho^{(n_R)}_{11}
-\rho^{(n_R)}_{22}] -\frac{\Gamma_L}{2}\rho^{(n_R)}_{12}\nl
&+\frac{\Gamma_L}{2}\rho^{(n_R)}_{34}+\Gamma_R\rho^{(n_R-1)}_{34}\\
%%%%%%%
\dot{\rho}^{(n_R)}_{33}=&i\Omega[\rho^{(n_R)}_{34}-\rho^{(n_R)}_{43}]
-(\Gamma_R+\Gamma_L)\rho^{(n_R)}_{33}\\
%%%%%%%%%%%%%%
\dot{\rho}^{(n_R)}_{44}=&i\Omega[\rho^{(n_R)}_{43}-\rho^{(n_R)}_{34}]
+\Gamma_L\rho^{(n_R)}_{22}-\Gamma_R \rho^{(n_R)}_{44}\\
%%%%%%%%%%%
\dot{\rho}^{(n_R)}_{34}=&-i(\epsilon+U)\rho^{(n_R)}_{34}+i\Omega
[\rho^{(n_R)}_{33}-\rho^{(n_R)}_{44}]\nl &+\frac{\Gamma_L}{2}
\rho^{(n_R)}_{12} - (\Gamma_R+\frac{\Gamma_L}{2}) \rho^{(n_R)}_{34}
\end{align}
\end{subequations}
Here, $\epsilon=E_a-E_b$ and $\Gamma_{L/R}=2\pi|\Omega_{L/R}|^2g_{L/R}$,
where $g_{L/R}$ is the density of states of the SET leads.
For simplicity, the assumption of the wide-band limit implies
$\Omega_{L/R}\equiv\Omega_{L/R k}$,
and makes $\Gamma_{L/R}$ energy independent.
In addition, low temperature conditions and $U\gg \Omega$ were assumed
to further simplify the equations.
Similarly, for model (II), we have \cite{AK}
\begin{subequations}
\begin{align}
\dot{\rho}^{(n_R)}_{11}=&i\Omega[\rho^{(n_R)}_{12}-\rho^{(n_R)}_{21}]
-\Gamma_L^\prime\rho^{(n_R)}_{11}+\Gamma_R^\prime\rho^{(n_R-1)}_{33}\\
%%%%%%%%%%
\dot{\rho}^{(n_R)}_{22}=&i\Omega[\rho^{(n_R)}_{21}-\rho^{(n_R)}_{12}]
-\Gamma_L\rho^{(n_R)}_{22}+\Gamma_R \rho^{(n_R-1)}_{44}
\\
%%%%%%%%%%%%%
\dot{\rho}^{(n_R)}_{12}=&-i\epsilon\rho^{(n_R)}_{12}+i\Omega[\rho^{(n_R)}_{11}
-\rho^{(n_R)}_{22}]\nl
&-\frac{\Gamma_L+\Gamma_L^\prime}{2}\rho^{(n_R)}_{12}+
\frac{\Gamma_R+\Gamma_R^\prime}{2}\rho^{(n_R-1)}_{34}\\
%%%%%%%
\dot{\rho}^{(n_R)}_{33}=&i\Omega[\rho^{(n_R)}_{34}-\rho^{(n_R)}_{43}]
+\Gamma_L^\prime\rho^{(n_R)}_{11}-\Gamma_R^\prime\rho^{(n_R)}_{33}\\
%%%%%%%%%%%%%%
\dot{\rho}^{(n_R)}_{44}=&i\Omega[\rho^{(n_R)}_{43}-\rho^{(n_R)}_{34}]
+\Gamma_L\rho^{(n_R)}_{22}-\Gamma_R \rho^{(n_R)}_{44}\\
%%%%%%%%%%%
\dot{\rho}^{(n_R)}_{34}=&-i(\epsilon+U)\rho^{(n_R)}_{34}+i\Omega
[\rho^{(n_R)}_{33}-\rho^{(n_R)}_{44}]\nl
&+\frac{\Gamma_L+\Gamma_L^\prime}{2}\rho^{(n_R)}_{12}-
\frac{\Gamma_R+\Gamma_R^\prime}{2}\rho^{(n_R)}_{34}
\end{align}
\end{subequations}
Except for the specific conditions of model (II),
the other parameters are the same as those for model (I) (as mentioned above).

By applying the $n$-ME formulation in Eqs.\ (\ref{MacD})-(\ref{SNw}),
we can straightforwardly calculate the output power spectrum
for both set-ups considered here.
As for qubit measurement using a QPC detector, the signal of the qubit
oscillations is manifested as a peak in the noise spectrum
at frequency $2\Omega$,
while the measurement effectiveness
is characterized by the SNR,
i.e., the {\it peak-to-pedestal} ratio.
We denote the noise pedestal by $S_p$, and obtain it
from $S(\omega\rightarrow \infty)$.
In Fig.\ 6, we show the dependence of the SNR
on the detector's configuration symmetries.

\begin{figure}[h]
\begin{center}
\includegraphics[width=7.5cm]{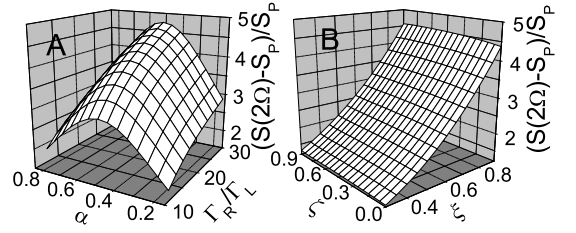}
\caption{\label{fig3} Signal-to-noise ratio: (A) for model (I), and
(B) for model (II).
%%%%
For model (I), we take $\Gamma_L = \Gamma$ as energy unit
and assume that $\mu_{L(R)}=\pm 50\Gamma$, $\Omega=2\Gamma$ and
$U=80\Gamma$.
%%%%%%%%%%%%
For model (II), we take $\bar{\Gamma}_L = \bar{\Gamma}$ as
energy unit and assume that $\Omega=\bar{\Gamma}$,
$U=50\bar{\Gamma}$, $\bar{\Gamma}_R=30\bar{\Gamma}$, and
$\alpha=\beta=1/2$. Zero temperature and $E_a=E_b$ are
assumed. }
\end{center}
\end{figure}
%%%%%%%
%%%%%%%%%%

%%%%%%%%%fig3

\begin{figure}[h]
\begin{center}
\includegraphics[width=7.5cm]{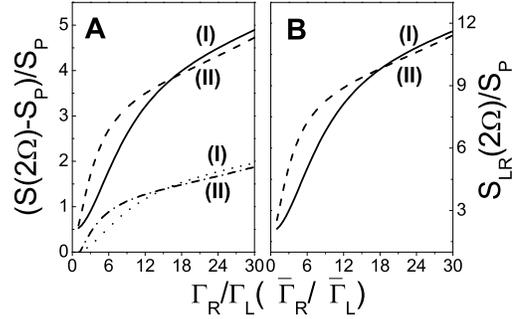}
\caption{Signal-to-noise ratio \emph{versus} tunnel-coupling
asymmetry, $\Gamma_R/\Gamma_L$ for model (I), and
$\bar{\Gamma}_R/\bar{\Gamma}_L$ for model (II). %%
In (A) the solid and dashed lines are the result in the presence of
cross correlation, while the dotted and dot-dashed lines are the
result after removing it. In (B) the mere cross correlation is
plotted. %%
$S_p$ is the pedestal noise of the entire circuit current.
$\xi=\zeta=0.9$, other parameters are the same as in Fig.\ 6. }
\end{center}
\end{figure}%%%%%%%
%%%%%%%%%%%%%%

%%%%%%%%fig4

\begin{figure}[h]
\begin{center}
\includegraphics[width=7.5cm]{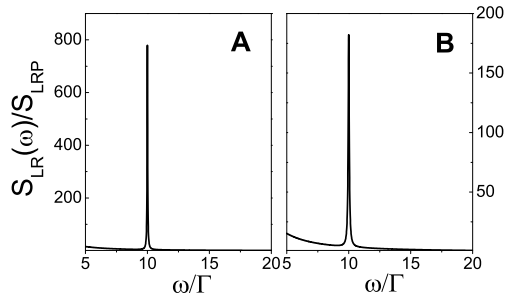}
\caption{ Spectral density of the cross-correlation scaled by its
own pedestal, which is defined here at twice the Rabi frequency of
the qubit oscillations. Parameters for model (I) in (A):
$\Gamma_L=0.05\Gamma$, $\Gamma_R=0.5\Gamma$, and $\Omega=5\Gamma$.
Parameters for model (II) in (B): $\bar{\Gamma}_L=0.05\Gamma$,
$\bar{\Gamma}_R=0.5\Gamma$, $\Omega=5\Gamma$, and $\xi=\zeta=0.9$.
In this plot we take $\Gamma$ as the energy unit.
Other conventions are the same as in Fig.\ 6. }
\end{center}
\end{figure}
%%%%%%%%%%%%%

The result of model (I) is shown in Fig.\ 6(A), where
we see that both the tunnel- and capacitive-coupling
symmetries crucially affect the measurement effectiveness.
The tunnel coupling asymmetry effect $\Gamma_R/\Gamma_L$ is due to the fact
that with the increase of $\Gamma_R/\Gamma_L$, the interaction time
between the detector electron and the qubit is decreased.
Thus, the detector's back-action is reduced and the SNR is enhanced
\cite{Gur05}.
For the effect of capacitive coupling, its degree of asymmetry
affects the contribution weight of the cross-correlation between
$I_L(t)$ and $I_R(t)$ to the entire circuit noise.
Specifically, the cross-correlation has a more important contribution for more
symmetric coupling, as shown in Fig.\ 6(A) by the $\alpha$-dependence.
This is because, as we shall demonstrate below,
the cross-correlation has a much higher {\it peak-to-pedestal} ratio
than the auto-correlation.

An unexpected feature observed in Fig.\ 6(A) is that under proper conditions,
i.e., for symmetric capacitive coupling and strongly asymmetric tunnel
coupling, the SNR can exceed ``4",
which is the upper bound quantum mechanically limited
on \emph{any linear response detectors} \cite{K-A01}.
However, whether this upper bound is applicable
to \emph{nonlinear} response detector is unclear {\it in priori},
since, in this case, the linear response relation between the current
and qubit state breaks down.
Hence, the subsequent Cauchy-Schwartz-inequality-based argument
leading to the upper bound of ``4" is not valid \cite{But06}.

To support the above theory, we further study model (II).
The result is presented in Fig.\ 6(B).
As explained in the model description, the parameters
$\xi$ and $\zeta$ characterize, respectively,
the left and right tunnel-coupling responses to the qubit states.
Fig.\ 6(B) shows an asymmetric tunnel coupling detector,
with $\gamma\equiv\bar{\Gamma}_R/\bar{\Gamma}_L=30$, which can
lead to higher SNR, because of the
weaker back-action from the detector, similar to that for model (I).
Here, we find that the SNR is insensitive to the right
junction response $\zeta$, but sensitive to the left one $\xi$.
Again, in this model, we observe that the SNR can violate the
K-A bound ``4" in the strong response regime.

We present further explanation
of the violation of the K-A bound.
Since $I(t)=\alpha I_L(t)+\beta I_R(t)$,
the current correlator $\la I(t) I(0) \ra$ contains the component
$S_{LR}(t)\equiv \la I_L(t) I_R(0)+I_R(t) I_L(0)\ra$,
i.e., the cross-correlation.
In addition, in the previous results, we see that for
more symmetric capacitive coupling, the SNR is larger, and
reaches a maximum at $\alpha=\beta=1/2$.
This feature indicates that the cross-correlation can enhance
the SNR.
Indeed, for the SET detector, both the left and right junction currents
($I_L$ and $I_R$) contain information of the qubit state; hence, their
``signal" parts are correlated.
This leads to a heuristic opinion that views the two junctions
as two detectors, like in the scheme of qubit measurement using two
point contacts proposed recently by Jordan and B\"uttiker \cite{But06},
where they found that the SNR of the cross-correlation can
strongly violate the K-A bound,
because of the negligibly small pedestal of the cross noise.
In our case, since $I_L(t)$ and $I_R(t)$ are subject to
a constraint from charge conservation,
the cross noise background of $I_L(t)$ and $I_R(t)$ does not vanish
in principle, unlike for the two independent QPC detectors \cite{But06}.
Nevertheless, the pedestal of the cross noise of the SET is much
smaller than that of the auto-correlation, which leads to an
enhanced SNR in the spectral density of the total circuit current
and to the violation of the K-A bound, as clearly shown in Fig.\ 7(A).
For comparative purposes, in Fig.\ 7(B), we plot the SNR of the cross-correlation, scaled by the noise pedestal $S_p$ of the circuit current.

In Fig.\ 8, the spectral density of the cross-correlation, scaled
by its own noise pedestal, is shown representatively.
As mentioned above, since the cross noise pedestal
is negligibly small at a high frequency limit,
we artificially (but more physically in some sense)
define the pedestal at a finite frequency, e.g.,
twice the qubit oscillation frequency. Obviously, the large SNR of the
cross-correlation drastically violates the K-A bound.
This result indicates that for qubit measurement using an SET,
one can explore the cross-correlation, rather than the auto-correlation,
as a probe of coherent oscillations.
In practice, such scheme is simpler
than the technique of QND measurement \cite{Ave02},
and holds the most advantages of SET over QPC.

% ========================================================
\section{Application to Quantum Transport}

{\flushleft We illustrate} the application of the $n$-ME approach
to quantum transport by first using a single-level quantum dot to
show the simple results of the $n$-resolved master equation and
then considering two more interesting examples.

\subsection{Single-Level Quantum Dot}

{\flushleft For transport through} a single-level ($E_0$)
quantum dot, under wide-band approximation,
the reservoir correlation functions can be expressed as
$ C_{\alpha}^{(\pm)}(t-\tau)=|t_{\alpha}|^2\sum_k
  e^{\pm i\ep_k(t-\tau)}
               n^{(\pm)}_{\alpha}(\ep_k) $,
where $n^{(+)}_{\alpha}(\ep_k)= n_{\alpha}(\ep_k)$
is the Fermi distribution function and
$n^{(-)}_{\alpha}(\ep_k)= 1-n_{\alpha}(\ep_k)$.
Then, the spectral functions are obtained as
\bea\label{C+-L}
A_{\al}^{(\pm)}=C_{\alpha}^{(\pm)}(\pm{\cal L})a
=\Gamma_{\alpha}n^{(\pm)}_{\alpha}(E_0)a \, .
\eea
Here, $\Gamma_{\alpha}=2\pi g_{\alpha}|t_{\alpha}|^2$,
where $g_{\alpha}$ is the density of states of the ``$\alpha$" lead.
%%%%
In the special case of zero temperature and a large bias voltage
$\mu_L\gg E_0\gg \mu_R$, we have
%%\bea\label{A+-LR}
$A^{(+)}_{L} = \Gamma_L a$, $A^{(-)}_{L} = 0$ \,,
$A^{(-)}_{R} = \Gamma_R a$, and $A^{(+)}_{R} = 0$ \,.
%%\eea
Substituting these into \Eq{ME-3a} yields
\bea\label{ME-RT}
\dot{\rho}^{(n)}
   &=&  -i {\cal L}\rho^{(n)} - \frac{1}{2}
   \left\{ [\Gamma_R a^{\dg}a\rho^{(n)}+\Gamma_L\rho^{(n)}aa^{\dg} \right. \nl
   & & \left. - \Gamma_L a^{\dg}\rho^{(n)}a
   -  \Gamma_R a\rho^{(n-1)}a^{\dg}]+{\rm H.c.} \right\}  \,.
\eea
Choosing the empty state $|0\ra$ and the occupied one
$|1\ra$, we obtain
\bea\label{rho-0011}
\dot{\rho}^{(n)}_{00}&=&-\Gamma_L\rho^{(n)}_{00}+\Gamma_R\rho^{(n-1)}_{11}\,, \nl
\dot{\rho}^{(n)}_{11}&=&-\Gamma_R\rho^{(n)}_{11}+\Gamma_L\rho^{(n)}_{00} \,.
\eea
This is the result derived by Gurvitz {\it et al}
under the limits mentioned above \cite{Gur96b}.
Applying the above $n$-ME, one can easily
perform all the transport studies outlined in Sec.\ II.

\subsection{Parallel Double Dots}

{\flushleft The system of two quantum dots}
coupled in parallel to two
reservoirs has attracted a great deal of attention as a realization
of a mesoscopic Aharonov--Bohm interferometer \cite{holl,chen,sigr}.
Indeed, such a system pierced by an external magnetic field,
as shown in Fig.\ 9, is an interference device whose transmission can
be tuned by varying the magnetic field. In the absence of the
interdot electron--electron interaction, the interference effects in
the resonant current through this system are quite transparent. This
is not the case, however, for interacting electrons \cite{gefen,neder}.

We consider a strong interdot electron-electron
repulsion---a Coulomb blockade. While the two dots may be
occupied simultaneously in the noninteracting
model, the Coulomb blockade prevents this. At first,
one might not expect that this repulsion could dramatically
modify the resonant current's dependence on the magnetic field.
We find, however, that the resonant current is completely blocked
for any value of the magnetic flux except for integer multiples of
the flux quantum ($\Phi_0=h/e$) \cite{DD-1}.
This striking effect goes far beyond our simple expectations.

Consider a double dot (DD) connected in parallel to two reservoirs,
as shown in Fig.\ 9. For simplicity, we consider spinless
electrons. We also assume that each of the dots contains only one
level, $E_{1}$ and $E_{2}$. In the presence of a
magnetic field, the system can be described by the following
tunneling Hamiltonian \cite{DD-1},
\begin{align}
H=H_0+H_T+\sum_{\mu =1,2} E_\mu d_\mu^\dagger d_\mu +Ud_1^\dagger
d_1 d_2^\dagger d_2\, . \label{a1}
\end{align} Here, the first term, $H_0=\sum_k [E_{kL}a_{kL}^\dagger a_{kL}
+E_{kR}a_{kR}^\dagger
a_{kR}]$, describes the reservoirs and $H_T$ describes their coupling
to the dots,
\begin{align}
H_T=\sum_{\mu,k}\Big (t_{\mu L}d_\mu^\dagger a_{kL} +t_{\mu
R}a_{kR}^\dagger d_\mu\Big )+ {\rm H.c.}\, , \label{a2}
\end{align}
where $\mu=1,2$ and $a_{kL}^\dagger$ and~$a_{kR}^\dagger$ are the
creation operators for the electrons in the reservoirs while
$d_{1,2}^\dagger$ is the creation operator for the DD. The last
term in Eq.~(\ref{a1}) describes the interdot repulsion. We assume
that there is no direct transmission between the dots and that the
couplings of the dots to the leads, $t_{\mu L(R)}$, are independent
of energy. In the absence of a magnetic field, one can always choose
the gauge in such a way that all couplings are real. In the presence
of a magnetic flux $\Phi$, however, the tunneling amplitudes between
the dots and the reservoirs are generally complex. We obtain $t_{\mu
L(R)}={\bar t}_{\mu L(R)}e^{i\phi_{\mu L(R)}}$, where $\bar t_{\mu
L(R)}$ is the coupling without the magnetic field. The phases around
the closed circle are constrained to satisfy
$\phi_{1L}+\phi_{1R}-\phi_{2L}-\phi_{2R}=\phi$,
where $\phi\equiv 2\pi\Phi/\Phi_0$.

\begin{figure}
\includegraphics[width=7cm]{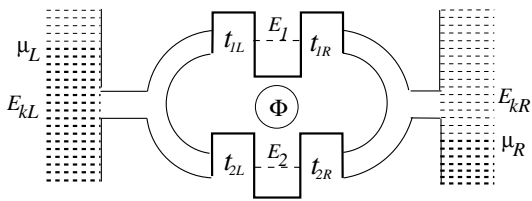}
\caption{Resonant tunneling through two parallel dots pierced by the
magnetic flux $\Phi$.} \label{fig1}
\end{figure}

Let the initial state of the system correspond to filling the left
and right reservoirs at zero temperature with electrons up to the
Fermi energies $\mu_L$ and $\mu_R$, respectively.
In the case of large bias, $|\mu_{L,R}-E_{1,2}|\gg\Gamma$,
applying either an exact single-particle wavefunction method
or the ME approach for non-interacting case,
we find a simple expression for the total current \cite{DD-1}
\begin{align} I(\phi ) =
I_0{\Delta^2+\Gamma_L\Gamma_R\sin^2 \phi
\over\Delta^2+\Gamma_L\Gamma_R\sin^2{\displaystyle\phi\over\displaystyle
2}\, }\, , \label{a6}
\end{align}
where $\Delta=E_1-E_2$ is the offset of the  dot levels,
and $I_0=2\Gamma_L\Gamma_R/\Gamma$ is the current for
non-interacting electrons in the absence of the magnetic filed,
with
$\Gamma_{L(R)}=2\pi D_{L(R)}|\bar{t}_{L(R)}|^2$
and $D_{L(R)}$ is the density of states of the leads.
The $\phi$-dependence in Eq.(\ref{a6})
is an example of the Aharonov--Bohm effect.

Next, we detail the study
for interacting dots in the Coulomb blockade case,
$E_{1,2}+U\gg \mu_L$, which excludes the states
corresponding to a simultaneous occupation of the two dots.
In this case, the Hilbert space of the DD state is reduced
to $|0\ra \equiv |00\ra$, $|1\ra \equiv |10\ra$, and $|2\ra \equiv |01\ra$,
where $|10\ra$ means the upper dot occupied and the lower dot unoccupied,
and other states have similar interpretations.
Applying \Eq{ME-3a}, we obtain \cite{DD-2}
\begin{widetext}
\begin{subequations}
\bea
\dot{\rho}^{(n)}_{00}=
-2 \Gamma_L \rho_{00}^{(n)}
+ \Gamma_R \rho_{11}^{(n-1)}
+ \Gamma_R\rho_{22}^{(n-1)}
+e^{i \left(\phi_{\text{R1}}-\phi_{\text{R2}}\right)}
\Gamma_R \rho_{12}^{(n-1)}
+e^{i \left(\phi_{\text{R2}}-\phi_{\text{R1}}\right)}
\Gamma_R \rho_{21}^{(n-1)}
\eea
%2222
\bea
\dot{\rho}_{11}^{(n)}=
\Gamma_L \rho_{00}^{(n)}
- \Gamma_R \rho_{11}^{(n)}
-\frac{1}{2} e^{i \left(\phi_{\text{R1}}-\phi_{\text{R2}}\right)}
\Gamma_R \rho_{12}^{(n)}
-\frac{1}{2} e^{i \left(\phi_{\text{R2}}-\phi_{\text{R1}}\right)}
\Gamma_R \rho_{21}^{(n)}
\eea
%3333
\bea
\dot{\rho}_{22}^{(n)}=
\Gamma_L \rho_{00}^{(n)}
- \Gamma_R\rho_{22}^{(n)}
-\frac{1}{2} e^{i \left(\phi_{\text{R1}}-\phi_{\text{R2}}\right)}
\Gamma_R \rho_{12}^{(n)}
-\frac{1}{2} e^{i\left(\phi_{\text{R2}}-\phi_{\text{R1}}\right)}
\Gamma_R \rho_{21}^{(n)}
\eea
%44444
\bea
\dot{\rho}_{12}^{(n)}=
e^{i \left(\phi_{\text{L1}}-\phi_{\text{L2}}\right)}
\Gamma_L \rho_{00}^{(n)}
-\frac{1}{2} e^{i \left(\phi_{\text{R2}}-\phi_{\text{R1}}\right)}
\Gamma_R \rho_{11}^{(n)}
-\frac{1}{2} e^{i \left(\phi_{\text{R2}}-\phi_{\text{R1}}\right)}
\Gamma_R \rho_{22}^{(n)}
- (\Gamma_R + i\Delta) \rho_{12}^{(n)}
\eea
%5555
\bea
\dot{\rho}_{21}^{(n)}=
e^{i \left(\phi_{\text{L2}}-\phi_{\text{L1}}\right)} \Gamma_L \rho_{00}^{(n)}
-\frac{1}{2} e^{i \left(\phi_{\text{R1}}-\phi_{\text{R2}}\right)}
\Gamma_R\rho_{11}^{(n)}
-\frac{1}{2} e^{i \left(\phi_{\text{R1}}-\phi_{\text{R2}}\right)}
\Gamma_R \rho_{22}^{(n)}
- (\Gamma_R-i\Delta) \rho_{21}^{(n)}
\eea
\end{subequations}
\end{widetext}
Owing to the neglected spin degrees of freedom
in constructing the Hilbert space, as a compensation,
here we have replaced $\Gamma_L$ with $2\Gamma_L$ to
equivalently restore its effect.

% \Delta

Applying the $n$-ME approach, we obtain the total current
in the steady-state limit \cite{DD-1,DD-2}
\begin{align}
I(\phi )=I_C{\Delta^2\over \Delta^2+I_C\left(2\Gamma_R
\sin^2{\displaystyle \phi\over\displaystyle 2} -\Delta\sin \phi
\right)}\, , \label{a10} \end{align}
where
$I_C=2\Gamma_L\Gamma_R/(2\Gamma_L+\Gamma_R)$ is the total current
(with the Coulomb blockade) in the absence of the magnetic field. Let
us compare Eq.~(\ref{a10}) with Eq.~(\ref{a6}) for the noninteracting
case. We find that for $\Delta \not =0$ both currents display
the Aharonov--Bohm oscillations.
However, their behavior is drastically different when $\Delta\to 0$.
The resonant current for the noninteracting electrons keeps
oscillating with the magnetic field, while in the case
of the Coulomb blockade, the current becomes non-analytic
in $\phi$. From
Eq.~(\ref{a10}), we can easily obtain that $I=I_C$ for $\phi =2\pi n$,
where $n=\Phi/\Phi_0$ is an integer, but $I=0$
for any other value of  $\Phi$.
Such an unexpected ``switching'' behavior of the electron current
in the magnetic field represents a non-trivial interplay of the
{\it Coulomb blockade} and {\it quantum interference}.

For understanding the switching phenomenon, it is desirable to
disentangle these two effects via a state basis transformation,
by defining new basis DD states,
$d^{\dagger}_\mu|0\rangle\to\tilde d^{\dagger}_\mu |0\rangle$,
chosen such that
$\tilde d^\dagger_2 |0\rangle$ is not coupled to the right
reservoir, {\em i.e.}, $t_{2R}\to \tilde t_{2R}=0$,
then the current
would flow only through the state $\tilde d_{1}^\dagger|0\rangle$.
This can be realized using the unitary transformation \cite{DD-1,DD-2}
\begin{align}\label{a11}
\left (\begin{array}{c}\widetilde {d}_{1}\\
\widetilde {d}_{2}\end{array}\right)={1\over {\cal
N}}\left(\begin{array}{cc}t_{1R}
& t_{2R}\\
-t^*_{2R}&t^*_{1R}\end{array}\right)\left (\begin{array}{c}d_{1}\\
d_{2}\end{array}\right),
\end{align}
where ${\cal N}=(\bar t_{1R}^2+\bar t_{2R}^2)^{1/2}$, which indeed
results in $\widetilde t_{2R}=0$.
In addition, the coupling of $\widetilde d^\dagger_2|0\rangle$
to the left lead reads
\begin{align}
\widetilde t_{2L}(\phi)=-e^{i(\phi_{2L}-\phi_{1R})}(\bar t_{1L}\bar
t_{2R}\, e^{i\phi}-\bar t_{2L}\bar t_{1R})/{\cal N} .\label{a12}
\end{align}
It follows from this expression that $\widetilde t_{2L}=0$ for
$\phi=2n\pi$ provided that
$\bar t_{1L}/\bar t_{2L}=\bar t_{1R}/\bar t_{2R}$, or for
$\phi=(2n+1)\pi$ if $\bar t_{1L}/\bar t_{2L}=-\bar t_{1R}/\bar t_{2R}$.
Obviously, for noninteracting DD,
$\widetilde d^\dagger_2|0\rangle$
has no contribution to current,
while $\widetilde d^\dagger_1|0\rangle$ carries a magnetic-flux
modulated current.
In the case of inter-dot Coulomb blockade,
however, the coupling of $\widetilde d^\dagger_2|0\rangle$
to the left lead is zero, which is of crucial importance.
If $\widetilde t_{2L}\neq 0$, then the state
$\widetilde d_1^\dagger |0\rangle$, carrying the current, will be
blocked by the inter-dot Coulomb repulsion.
As a result, the total current {\em vanishes}.
However, if the state $\widetilde d_2^\dagger |0\rangle$
is decoupled from {\em both\/} leads, it
remains unoccupied, so that the current can flow through the state
$\widetilde d_1^\dagger |0\rangle$. As shown above, this takes place
precisely for $\bar t_{1L}/\bar t_{2L}=\pm\,\bar t_{1R}/\bar
t_{2R}$. If this condition is not fulfilled, the current is always
zero, even for $\phi =2\pi n$.

Below we consider further the current fluctuations.
The shot-noise spectrum can be conveniently calculated
using the $n$-ME and the MacDonald's formula.
Noticeably,
for the present Coulomb blockade DD interferometer, we find that
the zero-frequency shot noise can be highly super-Poissonian,
and can even become divergent as $\Delta\rightarrow 0$.
For the coherent DD interferometer, analytical result of
the frequency-dependent noise can be obtained as \cite{DD-2}
\begin{widetext}
\begin{align}\label{S-R}
S(\omega)= \frac{8\Gamma_L\Gamma_R [ 2\Gamma_L\Gamma_R\Delta^2
    -\Delta^4+3\Delta^2\omega^2 -2\omega^2(\Gamma^2_R +\omega^2)]\bar{I}}
    {[(2\Gamma_L+\Gamma_R)\Delta^2 - (2\Gamma_L+3\Gamma_R)\omega^2]^2
      +\omega^2(2\Gamma_L\Gamma_R+2\Gamma^2_R+\Delta^2-\omega^2)^2}
    + 2\bar{I} ~.
\end{align}
\end{widetext}
Here, we have assumed $\phi=2\pi n$.
At zero frequency limit, the Fano factor can be given as
\begin{align}\label{F-1}
F\equiv\frac{S(0)}{2\bar{I}}
= \frac{  8\Gamma^2_L\Gamma^2_R + (4\Gamma^2_L+\Gamma^2_R) \Delta^2  }
{ (2\Gamma_L+\Gamma_R)^2 \Delta^2 }.
\end{align}
Noticeably, as $\Delta\rightarrow 0$, it becomes divergent!
Note that this divergence is not caused by the average current
$\bar{I}$, but by the zero-frequency noise itself.
In addition, from \Eq{S-R},
we find that the limiting order of
$\Delta\rightarrow 0$ and $\omega\rightarrow 0$,
would lead to different results i.e., if $\Delta\rightarrow 0$,
then $\omega\rightarrow 0$, the result can be given as
\begin{align}\label{F-2}
F=\frac{\Gamma^2_L+\Gamma^2_R}{(\Gamma_L+\Gamma_R)^2},
\end{align}
which is finite and coincides with the Fano factor of the single-level
transport \cite{Li053}.
The limiting order leading to \Eq{F-2}, which implies that we are
considering the noise for aligned DD levels.
In this case, as constructed above, see \Eq{a11} and Fig.\ 10(a),
the two transformed dot-states are decoupled to each other,
and one of them is also decoupled to both leads if $\phi=2\pi n$.
As a result, equivalently, the transport is through a single channel,
leading to the Fano factor \Eq{F-2}.

\begin{figure}
\center
\includegraphics[scale=0.7]{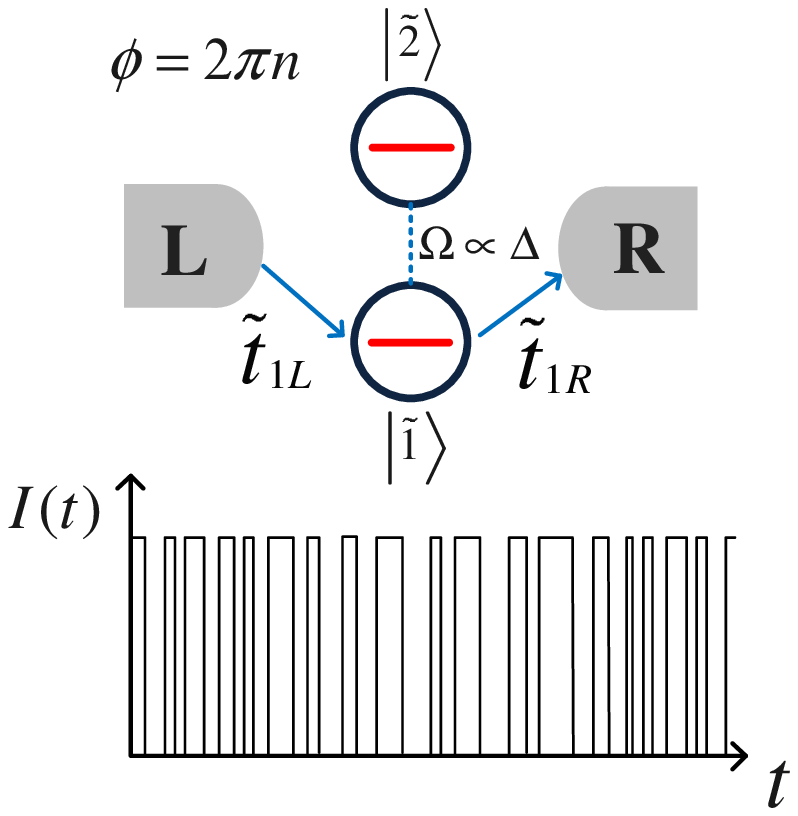}
\caption{
 Schematic interpretation for the noise divergence.
\emph{Upper panel}: Effective coupling of the DD to the leads and
between the dots, in the representation of transformed DD states, i.e.,
$|\ti{1}\ra\equiv\ti{d}^{\dg}_1  |0\ra$ and
$|\ti{2}\ra\equiv\ti{d}^{\dg}_2  |0\ra$.
\emph{Lower panel}: Coarse-grained temporal current, with a telegraphic
noise nature which causes divergence of the zero-frequency noise
when $\Delta\rightarrow 0$ . }
 \end{figure}

However, for $\Delta \rightarrow 0$ but $\neq 0$,
the situation is subtly different.
In this case, the two transformed states are weakly coupled,
with a strength $\propto\Delta$.
Thus, the transporting electron on state $\tilde{d}_1^{\dg}|0\ra$
can occasionally tunnel to $\tilde{d}_2^{\dg}|0\ra$,
which is disconnected to both leads,
and its occupation will block the current until the electron tunnels
back to $\tilde{d}_1^{\dg}|0\ra$ and arrives at the right lead.
Typically, this strong {\it bunching behavior}, induced by the interplay
of Coulomb interaction and quantum interference,
is well characterized by a profound super-Poissonian statistics.
In Fig.\ 10(b), the coarse-grained temporal current
with a telegraphic noise nature is plotted schematically.
We observe that, as $\Delta \rightarrow 0$, the current switching would become
extremely slow, leading to very long time ($\sim 1/\Delta$) correlation
between the transport electrons.
This long-time-scale fluctuation, or equivalently,
the low frequency component filtered out from the current,
which causes divergence of the shot noise as $\Delta \rightarrow 0$.
This is similar, in a certain sense, to the well known $1/f$ noise,
which goes to divergence as $f \rightarrow 0$.

\subsection{Probe of Majorana Fermion}

{\flushleft In this subsection, we apply the $n$-ME approach}
to analysis for a possible probe of the Majorana fermions \cite{MF-1,MF-2,MF-3}.
The Majorana fermions, proposed in 1937 by Majorana \cite{MF37},
are exotic particles since each Majorana fermion
is its own antiparticle \cite{Wil0910}.
The search for Majorana fermions in solid states,
as emerged quasiparticles (elementary excitations),
has been attracting a great deal of attention
\cite{Kit01,Fu08,Ore10,Sau10,Lut10,Ali10,Sau12}.
As a real example, an effective $p$-wave superconductor
can be realized using a semiconductor nanowire
with Rashba spin-orbit interaction and Zeeman splitting, and
in proximity to an $s$-wave superconductor \cite{Ore10,Sau10,Sau12}.
Of crucial importance is then a full experimental demonstration
of the Majorana fermion in solid states \cite{Kou12}.

Therefore, let us consider the system in Fig.\ 11.
The setup describes transport through a semiconductor quantum dot (QD),
while the QD is tunnel-coupled to a semiconductor nanowire
on an $s$-wave superconductor \cite{Kou12}.
It was found in Ref.\ \cite{Liu11} that
the Majorana bound state (MBS), emerged at the end of the nanowire,
will dramatically influence the {\it zero-bias} linear-response
conductance through the quantum dot,
as a result of the modified {\it static} spectral
property (the effective density-of-states) of the QD level.
%%%%
In the following section, going beyond linear response,
we consider transport through the quantum dot
under finite bias voltage and pay particular
attention to the Majorana's {\it dynamic} aspect \cite{MF-1,MF-2,MF-3}.
While a subtraction of the source and drain currents
can expose certain features of the Majorana fermion \cite{MF-1},
below we show that the more unique properties can be identified
from the shot noise \cite{MF-1,MF-2}, via a spectral dip
together with a pronounced zero-frequency noise enhancement effect.

\begin{figure}[!htbp]
  \centering
\includegraphics[width=5.5cm]{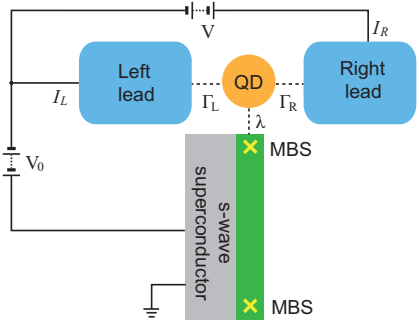}
\caption{
Schematic setup for detecting Majorana fermion
by transport through a semiconductor quantum dot (QD),
while the QD is coupled to a semiconductor nanowire
which is in contact with an $s$-wave superconductor.
Under appropriate conditions
a pair of Majorana bound states (MBS) is anticipated
to emerge at the ends of the nanowire. }\label{FIG1}
\end{figure}

Combining a strong Rashba spin-orbit interaction
and the Zeeman splitting, it was shown
that the proximity-effect-induced $s$-wave superconductivity in the nanowire
can support electron-hole quasiparticle excitations of Majorana bound states (MBS)
at the ends of the nanowire \cite{Ore10,Sau10,Sau12}.
Since the Zeeman splitting should be large enough in order to
drive the wire into a topological superconducting phase,
we can assume it much larger than the transport bias voltage,
the dot-wire coupling energy, and the dot tunneling rates
with the leads.
In this case, we can model the QD by a single resonant level
and treat the electron as spinless particle.
Accordingly, the entire system can be modeled using
$H=H_{\rm Leads}+H_{\rm sys}+H_T$.
$H_{\rm Leads}=\sum_{\alp=L,R}\sum_{k} \vep_{\alp k}\cdg_{\alp k}c_{\alp k}$
describes the normal metallic leads;
$H_T=\sum_{\alp=L,R}\sum_{k}t_{\alp k} d \cdg_{\alp k}+{\rm H.c.}$
is for the tunneling between the leads and the dot; and the
low-energy effective Hamiltonian for the central system is given as
\cite{MF-1,MF-2}
\bea\label{ham1}
    H_{\rm sys}=\eps_D d^{\dagger} d+\frac{i}{2}\eps_M\gamma_1\gamma_2
    +( \lambda d-\lambda^* d^{\dagger})\gamma_1 .
\eea
Here, $c^{\dagger}_{\alpha k}(c_{\alpha k})$ and $d^{\dagger}(d)$ are
the electron creation (annihilation) operators of the leads and the dot, respectively,
with corresponding energies of $\vep_{\alpha k}$ and $\epsilon_D$.
Particularly, in \Eq{ham1}, the second term describes the paired MBSs
generated at the ends of the nanowire and coupled to each other
by an energy $\epsilon_M\sim e^{-l/\xi}$, where $l$ is the wire length
and $\xi$ is the superconducting coherent length.
The last term in \Eq{ham1} describes the tunnel coupling between the dot
and the left MBS.
For spinless dot level, we can choose a real constant $\lambda$, while
in general $\lambda$ has a phase factor
associated with the spin direction.

To solve the transport problem associated with the Hamiltonian \Eq{ham1},
it is convenient to switch from the Majorana representation
to the regular fermion representation, through the exact transformation
$\gam_1=\fdg+f$ and $\gam_2=i(\fdg-f)$.
$f$ is the regular fermion operator, satisfying the
anti-commutative relation $\{f,\fdg\}=1$.
Accordingly, we rewrite $H_{\rm sys}$ as \cite{MF-1,MF-2}
\bea\label{ham2}
    H_{\rm sys}=\eps_D \ddg d+\eps_M\left(\fdg f-\frac{1}{2}\right)
    +\lam (d-\ddg)(\fdg+f)  .
\eea
For the convenience of the latter discussion, we rearrange the
tunnel coupling term in \Eq{ham2} as
$H_1=(\lam \fdg d + \lam_1 \fdg\ddg ) + {\rm H.c.}$,
where $\lam_1=\lam$ or $0$ corresponds to the dot
coupling to the MBS or to a regular fermion bound state.
In the transformed representation, the basis states of the central
system are given by $\ket{n_dn_f}$, where $n_d$ and $n_f$ can take the value of
0 or 1, so that we have four basis states
$\{\ket{00}, \ket{10}, \ket{01}, \ket{11}\}$.

Rather than the linear response \cite{Liu11},
we consider transport through the quantum dot
under {\it finite} bias voltage.
Associated with the voltage setup in Fig.\ 11,
the chemical potentials of the two leads are,
$\mu_L=eV_0$ and $\mu_R=e(V_0-V)$.
The so-called large bias limit indicates that
$|\mu_{L(R)}-\epsilon_D|$ is much larger than
the dot-level's broadening.
In this case, the dot level is deeply embedded into the voltage window
and the temperature effect is negligible in calculating
the transport currents. Moreover, this bias regime allows us to apply
the Born-Markov ME.
Its particle-number-resolved version is given as \cite{MF-1}
\beqn\label{rho1}
    \dot{\rho}^{(n)}&\!\!=\!\!&-i{\cal L} \rho^{(n)}
    \!-\!\frac{\Gam_L}{2}\left( d \ddg\rho^{(n)}
      +\rho^{(n)} d \ddg-2\ddg\rho^{(n)}d \right)\re
    &&-\frac{\Gam_R}{2}\left(\ddg d \rho^{(n)}
      +\rho^{(n)}\ddg d-2d\rho^{(n-1)}\ddg\right) ,
\eeqn
where ``$n$" represents the electron number transferred
through the central system, and $\rho^{(n)}$ satisfies the condition
$\sum_{n=0}^{\infty}\rho^{(n)}(t)=\rho(t)$.
Here, we introduced the Liouvillian superoperator as
${\cal L} \rho\equiv [H_{\rm sys}, \rho]$,
and the tunneling rate as
$\Gam_\alp=2\pi g_\alp|t_\alp|^2$,
where $g_\alp$ is the density-of-states of the lead $\alp$ ($L$ or $R$).
Corresponding to \Eq{rho1}, the {\it unconditional} Lindblad
ME is given as
$ \dot{\rho} = -i{\cal L} \rho + \Gamma_L {\cal D}[d^{\dagger}]\rho
 + \Gamma_R {\cal D}[d]\rho $,
where ${\cal D}[A]\rho\equiv A\rho A^{\dagger}
-\frac{1}{2}\{A^{\dagger}A, \rho\}$.

{\it Shot Noise.}---
Below we focus our interest on the shot noise spectrum, which
beyond the steady-state current,
reflects the {\it dynamic} aspect of the central system.
Specifically, we consider the current correlator
$S_\alp(t)=\frac{1}{2}\langle \{\delta I_\alp(t),
\delta I_\alp(0)\}\rangle$, where $\alp=L$ and $R$
and $\delta I_\alp(t)=I_\alp(t)-I^s_\alp$.
Here, $I^s_\alp$ is the steady-state current in the $\alp$-th lead,
and the (quantum statistical) average of defining the correlator
is over the steady state.
The shot noise spectrum $S_\alp(\omega)$,
i.e., the Fourier transform of $S_\alp(t)$,
can be calculated most conveniently
within the $n$-ME formalism using the McDonald's formula.

\begin{figure}[!htbp]
  \centering
\includegraphics[width=6.5cm]{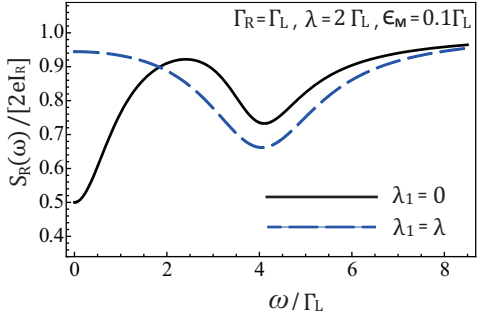}
  \caption{
Shot noise which reveals a spectral {\it dip}
and zero-frequency {\it enhancement} effect.
The former behavior reflects existence of
coherent oscillations
which indicates the formation of a bound state coupled
to the quantum dot, and the latter originates from
the nature of Majorana excitation.
``$\lambda_1=\lambda$" corresponds to the Majorana bound state
and ``$\lambda_1=0$" to a regular one.
As in Fig.\ 11, we set $\eps_D=0$ in this plot.  }
\end{figure}

Figure 12 displays the representative result for the shot noise spectrum.
First, we notice that a spectral {\it dip} appears
at the frequency $\omega_c\simeq 2\lam$ (for $\eps_D=0$ and small $\eps_M$),
which reflects an existence
of coherent oscillations in the central system,
where is $\omega_c$ the characteristic frequency.
This is an important signature, since it indicates
the emergence of {\it bound states} at the ends of the wire
with discrete energy, gapped from other higher energy continuum.
Here, we should note that the quantum dot is coupled to a
nanowire, and hence, in the usual case, the wire states are extended and have
continuous energies, which cannot support coherent oscillations
as indicated by the spectral ``dip" in Fig.\ 12.
As a comparison, in Fig.\ 12, we also plot the result of
the same quantum dot coupling to a regular bound state
with the same strength ($\lambda$).
While an ``oscillation dip" appears similarly at the same frequency,
nevertheless, the zero (and low) frequency noise
differs remarkably from the Majorana case.

To characterize the zero frequency noise, we use
the well-known Fano Factor, $F_\alp=S_\alp(0)/2eI_\alp^s$.
For symmetric rates $\Gam_R=\Gam_L=\Gam_0$,
we find $F_L=F_R$ and denote the Fano factor simply by $F$.
Analytically, we obtain \cite{MF-1}
\beq\label{DFF}
    F-F^{\rm (R)}\!=\frac{2\lam^2\lam_1^2}
    {(\Gam_0^2+\eps_M^2)(\lam^2+\lam_1^2)+4\lam^2\lam_1^2} .
\eeq
Here, $F^{\rm (R)}$ is the Fano factor of
coupling to a regular (R) bound state.
Moreover, in this result, we distinguish the coupling amplitudes
$\lam$ and $\lam_1$
(introduced in the coupling Hamiltonian).
We notice that $\lam$ and $\lam_1$
play identical (symmetric) roles and
the difference, \Eq{DFF}, vanishes
if any of the amplitudes disappears.
We then understand that the zero-frequency noise enhancement
is arising from the peculiar nature of the Majorana excitation.
Therefore, the noise enhancement effect in Fig.\ 12
is another useful signature for Majorana excitation
at the ends of the nanowire.
In addition, \Eq{DFF} can be used to obtain the important parameters
of the Majorana's mutual interaction ($\eps_M$)
and its coupling to the quantum dot ($\lam$).
In particular, based on \Eq{DFF},
after setting $\lam_1=\lam$ for the Majorana fermion,
an even simpler result can be
obtained in the limit $\eps_M\rightarrow 0$ i.e., $F=1-\frac{1}{2}[1+2(\lam/\Gam_0)^2]^{-1}$.
This result provides a very simple relation between the
Fano factor and the scaled coupling amplitude ($\lam/\Gam_0$).

Now, we consider $\Gam_L\neq \Gam_R$.
Taking the limit $\eps_M\rightarrow 0$ and setting $\eps_D=0$,
we obtain \cite{MF-1}
\bsub
\beq
    F_R=\frac{\Gam_L^2+\Gam_R^2}{\Gam^2}
    +\frac{8\Gam_R\Gam_L\lam^2\left[(5-3y)\Gam^2+16\lam^2\right]}
    {\Gam^2(\Gam^2+8\lam^2)^2}  ,
\eeq
\beq
    F_L=\frac{\Gam_L^2+\Gam_R^2}{\Gam^2}
    +\frac{8\Gam_R\Gam_L\lam^2\left[(5-3/y)\Gam^2+16\lam^2\right]}
    {\Gam^2(\Gam^2+8\lam^2)^2}  ,
\eeq
\esub
We notice that the first term in $F_{L(R)}$ is the Fano factor
corresponding to transport through an isolated
single-level quantum dot \cite{Li053,ChT91},
while the second term arises from coupling to the Majorana fermion.
If $\Gam_L\neq \Gam_R$, we find that
$F_R-F_L=24(\Gam_L-\Gam_R)\Gam\lam^2/(\Gam^2+8\lam^2)^2\neq 0$.
(This difference vanishes when $\Gam_L=\Gam_R$.)
Finally, we mention that the steady-state current
through the quantum dot cannot reveal the Majorana information
in the symmetric case ($\Gam_L=\Gam_R$) \cite{MF-1}.
However, the Fano factor given above carries such information,
since the second term in $F_{L(R)}$ does not vanish when $y=1$.
Note also that, if the quantum dot couples to a regular bound state
(another dot), the zero frequency noise (Fano factor) is the same
as the first term of the above $F_{L(R)}$,
being unaffected by the side coupling.

\section{Master Equation under Self-consistent
         Born Approximation}

{\flushleft In this section, }
we review our recent work on improving the ME beyond
the second-order Born approximation \cite{SCBA-1,SCBA-2,SCBA-3}.
The basic idea is to base the formulation on the
{\it self-consistent} Born approximation.
That is, we replace the {\it free} Green's function
in the tunneling self-energy diagram by an {\it effective}
reduced propagator under the Born approximation \cite{SCBA-1}.
Remarkably,
we will see that this modification can greatly improve the results.

In some cases (such as in quantum optics), the second-order
ME works perfectly well.
However, for quantum transport, the second-order expansion of the
tunneling Hamiltonian only corresponds to sequential transport,
which does not incorporate the level broadening effect \cite{Li05b},
implying thus a validity condition of {\it large bias voltage}.
Moreover, for interacting systems,
although the second-order ME
can predict the Coulomb staircase behavior,
it cannot deal with the cotunneling and Kondo effects.
To overcome this limitation, higher-order expansions
of the tunneling Hamiltonian are required
\cite{Sch94,Sch96,Sch06,Yan080911,Yan2012,Wac05+10,CS11,
Leeu09,Galp09,Galp10,KG13}.

The second-order ME is obtained from
the well-known Born approximation through
perturbative expansion of the tunneling Hamiltonian \cite{Li05b}.
The resultant {\it dissipation} term, in analogy to
the quantum dissipative system,
corresponds to a {\it self-energy} process of tunneling.
On the other hand, it is well known that in the Green's function theory,
an efficient scheme of higher-order correction
is the use of renormalized self-energy diagram
under the {\it self-consistent Born} approximation (SCBA),
which is actually a type of self-consistent renormalization
to the bare propagator with a {\it dressed} one \cite{Mat}.
From this insight, for quantum transport
we may replace the free (system-Hamiltonian only) Green's function
in the second-order self-energy diagram, with an {\it effective}
propagator defined by the second-order ME \cite{SCBA-1}.
We will see that the effect of this improvement is remarkable:
it recovers not only the {\it exact} result
of noninteracting transport under {\it arbitrary} voltages
but also the {\it cotunneling} and nonequilibrium {\it Kondo}
features for interacting systems.

\subsection{Formulation of the SCBA-ME}

\subsubsection{Master equation under Born approximation}

{\flushleft In a more compact form}, we reexpress the ME
(\ref{ME-3a}) (under the second-order Born approximation) as
\cite{SCBA-1,SCBA-2,SCBA-3}
\begin{align}\label{BA-ME}
  \dot\rho(t) &=-i{\cal L}\rho(t)
  - \sum_{\mu\sigma}\Big\{\big[a^{\bar\sigma}_\mu,
  A^{(\sigma)}_{\mu\rho}(t)\big]
  +{\rm H.c.}      \Big\} .
\end{align}
Here, we introduced:
$\sigma=+$ and $-$, $\bar{\sigma}=-\sigma$;
$a^+_{\mu}=a^{\dagger}_{\mu}$, and $a^-_{\mu}=a_{\mu}$.
The superoperators can be expressed as
${\cal L}\rho=[H_S,\rho]$, and
$ A^{(\sigma)}_{\mu\rho}(t)
= \sum_{\alpha=L,R} A^{(\sigma)}_{\alpha\mu\rho}(t)$
while
%%\begin{align}\label{Arho-SCBA}
$ A^{(\sigma)}_{\alpha\mu\rho}(t)
 =\sum_\nu\int^t_0 d\tau C^{(\sigma)}_{\alpha\mu\nu}(t-\tau)
\left\{{\cal G}(t,\tau)[a^{\sigma}_\nu\rho(\tau)]\right\} $.
${\cal G}(t,\tau)$ is the free propagator, determined by the
{\it system} Hamiltonian as
${\cal G}(t,\tau)=e^{-i{\cal L}(t-\tau)}$.

Now we present a specific characterization
for $ C^{(\sigma)}_{\alpha\mu\nu}(t-\tau)$
in terms of its Fourier transform:
\begin{align}
C^{(\pm)}_{\alpha\mu\nu}(t-\tau)=\int^\infty_{-\infty}
\frac{d\omega}{2\pi}
e^{\pm i\omega (t-\tau)}\Gamma^{(\pm)}_{\alpha\mu\nu}(\omega).
\end{align}
Accordingly, we have
$\Gamma^{(+)}_{\alpha\mu\nu}(\omega)
=\Gamma_{\alpha\nu\mu}(\omega)n^{(+)}_{\alpha}(\omega)$
and $\Gamma^{(-)}_{\alpha\mu\nu}(\omega)
=\Gamma_{\alpha\mu\nu}(\omega)n^{(-)}_{\alpha}(\omega)$,
where $\Gamma_{\alpha\mu\nu}(\omega)
=2\pi\sum_{k}t_{\alpha\mu k}t^\ast_{\alpha\nu k}\delta(\omega-\epsilon_k)$
is the spectral density function of the reservoir ($\alpha$),
$n^{(+)}_{\alpha}(\omega)$ denotes the Fermi function $n_{\alpha}(\omega)$,
and $n^{(-)}_{\alpha}(\omega)=1-n_{\alpha}(\omega)$ is introduced for brevity.
Alternatively, we may introduce the Laplace transform
of $ C^{(\sigma)}_{\alpha\mu\nu}(t-\tau)$, denoted by
$C^{(\sigma)}_{\alpha\mu \nu}(\omega)$,
which is related to $\Gamma^{(\pm)}_{\alpha\mu \nu}(\omega)$
through the well known dispersive relation:
\begin{align}\label{FDT2}
C^{(\pm)}_{\alpha\mu \nu}(\omega)
&=\int^\infty_{-\infty}\frac{d\omega'}{2\pi}
\frac{i}{\omega\pm\omega'+i0^+}\Gamma^{(\pm)}_{\alpha\mu \nu}(\omega').
\end{align}

For the reservoir spectral density function,
we assume a Lorentzian form as
\be\label{Gammaw}
\Gamma_{\alpha\mu \nu}(\omega)=
\frac{\Gamma_{\alpha\mu \nu} W^2_\alpha}{(\omega-\mu_\alpha)^2+W^2_\alpha} .
\ee
Here, we use the constant $\Gamma_{\alpha\mu \nu}$
(without the argument $\omega$)
to denote the height of the Lorentzian spectrum,
and $W_\alpha$ to characterize its bandwidth.
The form of \Eq{Gammaw} also corresponds to a half-occupied band
for each lead, which peaks the Lorentzian center
at the chemical potential $\mu_\alpha$ of the lead.
Obviously, the usual constant spectral density function is
obtained from \Eq{Gammaw} in the limit $W_\alpha\rightarrow\infty$,
yielding $\Gamma_{\alpha\mu \nu}(\omega)=\Gamma_{\alpha\mu \nu}$.
Corresponding to the Lorentzian spectral density function,
straightforwardly, we obtain
\begin{align}
C^{(\pm)}_{\alpha\mu \nu}(\omega)
&=\frac{1}{2}\left[\Gamma^{(\pm)}_{\alpha\mu \nu}(\mp \omega)
 +i\Lambda^{(\pm)}_{\alpha\mu \nu}(\mp \omega)\right].
\end{align}
The imaginary part, through the dispersive relation,
is associated with the real part as
\begin{align}
& \Lambda^{(\pm)}_{\alpha\mu \nu} (\omega)
={\cal P}\int^\infty_{-\infty}\frac{d\omega'}{2\pi}
\frac{1}{\omega\pm\omega'}\Gamma^{(\pm)}_{\alpha\mu \nu}(\omega)
\nl&=\frac{\Gamma_{\alpha\mu\nu}}{\pi}
\Bigg\{{\rm Re}\left[\Psi\left(\frac{1}{2}
+i\frac{\beta(\omega-\mu_\alpha)}{2\pi}\right)\right]
\nla
-\Psi\left(\frac{1}{2}+\frac{\beta W_\alpha}{2\pi}\right)
\mp\pi\frac{\omega-\mu_\alpha}{W_\alpha}\Bigg\},
\end{align}
where ${\cal P}$ represents the principle value,
$\Psi(x)$ is the digamma function, and
$\beta=1/k_B T$ denotes the inverse temperature.

The second-order ME
can be applied only to transport
under large bias voltage i.e., the Fermi levels of the leads
should be considerably away from the system levels,
by at least several times of the level's broadening.

\subsubsection{Master equation under self-consistent Born approximation}

%% second-ME   second-order ME

{\flushleft The basic idea }
to improve the second-order ME can follow what is typically done
in the Green's function theory, i.e., correcting the self-energy diagram
from the {\it Born} to a {\it self-consistent Born} approximation.
In our case, the SCBA scheme can be implemented by
replacing the {\it free} propagator
in the second-order ME,
${\cal G}(t,\tau)=e^{-i{\cal L}(t-\tau)}$,
by an effective one, ${\cal U}(t,\tau)$,
which propagates a state
with the precision of the second-order Born approximation,
given by \Eq{BA-ME}.
From this consideration, the generalized SCBA-ME
follows \Eq{BA-ME} directly as \cite{SCBA-1}
\begin{align}\label{SCBA-ME}
  \dot\rho(t) &=-i{\cal L}\rho(t)
  - \sum_{\mu\sigma}\Big\{\big[a^{\bar\sigma}_\mu,
  {\cal A}^{(\sigma)}_{\mu\rho}(t)\big]
  +{\rm H.c.}      \Big\} .
\end{align}
Here, $ {\cal A}^{(\sigma)}_{\mu\rho}(t)
= \sum_{\alpha=L,R} {\cal A}^{(\sigma)}_{\alpha\mu\rho}(t)$,
and $ {\cal A}^{(\sigma)}_{\alpha\mu\rho}(t)
 =\sum_\nu\int^t_0 d\tau C^{(\sigma)}_{\alpha\mu\nu}(t-\tau)
\left\{{\cal U}(t,\tau)[a^{\sigma}_\nu\rho(\tau)]\right\} $.
To close this ME, let us define
$\ti\rho_j(t)\equiv {\cal U}(t,\tau)[a^{\sigma}_\nu\rho(\tau)]$
(here and in the following equation we use ``$j$" to denote the double
indices $(\nu,\sigma)$ for the sake of brevity).
Then, the equation-of-motion (EOM)
of this auxiliary object is given as \cite{SCBA-1}
\begin{align}\label{tirhoj}
  \dot{\ti\rho}_j(t) = -i{\cal L}\ti\rho_j(t)
-\int^t_{\tau}dt'  \Sigma^{(A)}_2(t-t')\ti\rho_j(t').
\end{align}
In this equation, we introduce a notation $\Sigma^{(A)}_2$
for the second-order self-energy superoperator,
where the superscript ``$(A)$" indicates an essential difference
from the usual one
because it involves {\it anticommutators}, rather than
the {\it commutators} in the second-order ME.
More explicitly, we have \cite{SCBA-1}
\begin{align}\label{Acmt}
\int^t_{\tau}dt' \Sigma^{(A)}_2(t-t')
& \ti\rho_j(t')= \sum_{\mu} \Big[\big\{a_\mu,A^{(+ )}_{\mu\ti\rho_j}\big\}
+\big\{a^\dg_\mu,A^{(-)}_{\mu\ti\rho_j}\big\}   \nl
& +\big\{a^\dg_\mu,A^{(+ )\dg}_{\mu\ti\rho_j}\big\}
+\big\{a_\mu,A^{(- )\dg}_{\mu\ti\rho_j}  \big\} \Big] ,
\end{align}
where $A^{(\pm)}_{\mu\ti\rho_j}$ is defined as
$ A^{(\sigma')}_{\mu\ti\rho_j}=\sum_{\alpha=L,R}\sum_{\nu'}
 \int^t_\tau dt' C^{(\sigma')}_{\alpha\mu\nu'}(t-t')
\left\{ e^{-i{\cal L}(t-t')}[a^{\sigma'}_{\nu'}
\tilde{\rho}_j(t')]\right\}$.
Because of the anticommutative brackets that appearing in \Eq{Acmt},
we stress that the propagation of $\ti\rho_j(t)$
is not governed by the usual second-order ME.
This, in certain sense, violates the
celebrated quantum regression theorem.
We notice that the second-order reduced propagator ${\cal U}$
was introduced from $\rho(t)={\cal U}(t,t_0)\rho(t_0)$,
and in \Eq{tirhoj}, the quantity being propagated is
$a^{\sigma}_\nu\rho(t_0)$,
which differs from the former only by an initial condition.
Then, from experience, we may expect that
the propagator must be independent of the initial condition,
in the present context, which is the object to be propagated.
In most cases, this statement is true.
However, our analysis shows that this general rule
(the celebrated quantum regression theorem),
quite unexpectedly, is not followed in the present case.
The basic reason is that the object being propagated,
$a^{\sigma}_\nu\rho$, contains an extra electron operator.
Owing to the Pauli principle (or Fermi-Dirac statistics),
extra negative signs appear in two of the four
self-energy terms in its equation-of-motion.
This changes the commutators in the usual master equation
to the anti-commutators in \Eq{Acmt}.
We find that this subtle issue is extremely important
-- otherwise we cannot obtain the correct results
such as the illustrative examples in this work.

\subsubsection{Steady state current}
%%\vspace{5cm}

{\flushleft Within the framework of SCBA-ME},
similar to its second-order counterpart,
the current through the $\alpha$th lead is given as \cite{SCBA-1}
\begin{align}
   I_{\alpha}(t)= 2\sum_\mu {\rm Re}
\left\{ {\rm Tr}\big[ {\cal A}^{(+)}_{\alpha\mu\rho}(t)a_\mu
  - {\cal A}^{(-)}_{\alpha\mu\rho}(t)a^\dg_\mu\big] \right\} .
\end{align}
Here, ${\rm Re}\{\cdots\}$ indicates the real part of $\{\cdots\}$
and ${\rm Tr}[\cdots]$ is the trace of $[\cdots]$
over the {\it system} only states.
For steady state,
consider the integral $\int^t_0 d\tau [\cdots]\rho(\tau)$
in ${\cal A}^{(\pm)}_{\alpha\mu\rho}(t)$.
Since physically, the correlation function
$C^{(\pm)}_{\alpha\mu\nu}(t-\tau)$ in the integrand
is nonzero only on {\it finite} timescale,
we can replace $\rho(\tau)$
in the integrand by the steady state $\bar{\rho}$,
in the long time limit ($t\rightarrow\infty$).
After this replacement, we obtain
\begin{align}
{\cal A}^{(\pm)}_{\alpha\mu\bar\rho} %(t\rightarrow\infty)
=\sum_\nu\int^\infty_{-\infty}\frac{d\omega}{2\pi}\,
\Gamma^{(\pm)}_{\alpha\mu\nu}(\omega)
{\cal U}(\pm\omega)[a^{\pm}_\nu\bar\rho] .
\end{align}
Then, substituting this result into \Eq{SCBA-ME},
we can directly solve for $\bar\rho$,
and calculate the steady state current.

Based on $\bar{\rho}$, in order to further obtain the current,
we first introduce
$\varphi_{1\mu\nu}(\omega)={\rm Tr} \big[a_\mu\ti\rho_{1\nu}(\omega)\big]$
and
$\varphi_{2\mu\nu}(\omega)={\rm Tr} \big[a_\mu\ti\rho_{2\nu}(\omega)\big]$,
where $\ti\rho_{1\nu}(\omega)$ and $\ti\rho_{2\nu}(\omega)$
are calculated using \Eq{tirhoj},
with an initial condition of
$\ti\rho_{1\nu}(0)=\bar\rho a^\dg_\nu$
and $\ti\rho_{2\nu}(0)=a^\dg_\nu\bar\rho$.
To simplify the notations, we denote the various matrices
expanded in the system state basis $\{ |\mu\ra,|\nu\ra \}$
in terms of a boldface form: $\bm\varphi_1(\omega)$,
$\bm\varphi_2(\omega)$, and $\bm{\Gamma}_{L(R)}$.
If $\bm\Gamma_L$ is proportional to $\bm\Gamma_R$ by a constant,
the steady state current can be recast
to the Landauer-B\"uttiker type \cite{SCBA-1}
\begin{align}
\bar I =  2~ {\rm Re}
\int^\infty_{-\infty}\frac{d\omega}{2\pi}
\left[ n_L(\omega)- n_R(\omega)\right] {\cal T}(\omega) ,
\end{align}
where the tunneling coefficient, very compactly, is given by
\begin{align}
 {\cal T}(\omega)={\rm Tr}\{
  \bm\Gamma_L\bm\Gamma_R
  (\bm\Gamma_L+\bm\Gamma_R)^{-1}
  {\rm Re}\big[\bm\varphi(\omega)\big]\}.
\end{align}
Here, $\bm\varphi(\omega)=\bm\varphi_1(\omega)+\bm\varphi_2(\omega)$.

Now, we demonstrate that for a noninteracting system,
the above stationary current
coincides precisely with the nonequilibrium Green's function approach,
both giving the exact result under a arbitrary bias voltage.
In general, a noninteracting system can be described by
$H_S=\sum_{\mu\nu}h_{\mu\nu}a^\dg_\mu a_\nu$.
We directly obtain the EOM of $\bm \varphi_i$ as follows:
\begin{align}\label{varphiw0}
-i\omega\bm\varphi_i(\omega)-\bm\varphi_i(0)=-i\bm h\bm\varphi_i(\omega)
-i\bm\Sigma_{0}(\omega)\bm\varphi_i(\omega).
\end{align}
$\bm\varphi_i(0)$ denotes the initial conditions for
$\varphi_{1\mu\nu}(0)={\rm Tr}\big[a_\mu\bar\rho a^\dg_{\nu}\big]$
and $\varphi_{2\mu\nu}(0)={\rm Tr}\big[a_\mu a^\dg_{\nu}\bar\rho\big]$.
The tunnel-coupling self-energy $\bm\Sigma_{0}$ is given by
$ \Sigma_{0\mu\nu}(\omega)
= -i\sum_{\alpha} \big[ C^{(-)}_{\alpha\mu\nu}(\omega)
+ C^{(+)\ast}_{\alpha\mu\nu}(-\omega)\big]$, or
\begin{align}\label{Self1}
\Sigma_{0\mu\nu}(\omega)
&=\int^\infty_{-\infty} \frac{d\omega'}{2\pi}
\frac{\Gamma_{\mu\nu}(\omega')}{\omega-\omega' +i0^+} .
\end{align}
Then, based on \Eq{varphiw0}, summing up $\bm\varphi_1(\omega)$
and $\bm\varphi_2(\omega)$ we obtain
%\be\label{phiomega}
%\bm\varphi(\omega)
%=\frac{i}{\bm h-\omega-\bm\Sigma_{0}(\omega)}=i\bm G^r(\omega). ~~ ?? \nonumber
%\ee
%%%
\be\label{phiomega}
\bm\varphi(\omega)
=i\big[ \omega-\bm h-\bm\Sigma_{0}(\omega)\big]^{-1}
%%=i\bm G^r(\omega).
\ee
For deriving this result, the cyclic property under trace
and the anti-commutator,
$\{a_\mu,a^\dg_\nu\}=\delta_{\mu\nu}$, have been used.
\Eq{phiomega} is nothing but the exact Green's function
for transport through a noninteracting system,
thus giving the exact stationary current after inserting it
into the above current formula.

\subsubsection{Interacting case}
%%\vspace{5cm}

{\flushleft To show the application of the proposed SCBA-ME }
in interacting systems, as an illustrative example,
we consider the transport through
an interacting quantum dot described as:
\be\label{Ands-H}
H_S = \sum_{\mu}\left(\epsilon_{\mu} a_{\mu}^{\dg}a_{\mu}
       +\frac{U}{2}n_{\mu}n_{\bar{\mu}}\right) .
\ee
Here, the index $\mu$ labels the spin up (``$\uparrow$") and spin
down (``$\downarrow$") states, and $\bar{\mu}$ represents the
opposite spin orientation.
$\epsilon_{\mu}$ denotes the spin-dependent energy level,
which may account for the Zeeman splitting in the presence of a
magnetic field ($B$),
$\epsilon_{\uparrow,\downarrow}=\epsilon_0\pm g\mu_B B$.
Here, $\epsilon_0$ is the degenerate dot level
in the absence of a magnetic field;
$g$ and $\mu_B$ are the Lande-$g$ factor and the Bohr's magneton, respectively.
In the interaction part,
$Un_{\uparrow}n_{\downarrow}$ is the Hubbard term,
$n_{\mu}=a^{\dg}_{\mu}a_{\mu}$ is the number operator,
and $U$ represents the interacting strength.

First, we notice that $C^{(\pm)}_{\alpha\mu\nu}$
is diagonal with respect to the spin states, i.e.,
$C^{(\pm)}_{\alpha\mu\nu}(t)=\delta_{\mu\nu}C^{(\pm)}_{\alpha\mu}(t)$
and
$\Gamma^{(\pm)}_{\alpha\mu\nu}=\delta_{\mu\nu}\Gamma^{(\pm)}_{\alpha\mu}$.
Here, $\delta_{\mu\nu}$ is the usual $\delta$-function with discrete
indices, and in $C^{(\pm)}_{\alpha\mu}$ and $\Gamma^{(\pm)}_{\alpha\mu}$,
there is only a single state index ($\mu$) for brevity.
Then, we specify the states involved in the transport as
$|0\ra$, $|\up\ra$, $|\down\ra$, and $|d\ra$, corresponding to
the empty, spin-up, spin-down, and double occupancy states, respectively,
Using this basis, we reexpress the electron operator
in terms of projection operator,
$a^\dg_\mu=|\mu\ra\la 0|+(-1)^{\mu}|d\ra\la \bar \mu|$,
where the conventions $(-1)^\up=1$ and $(-1)^\down=-1$ are implied.
For a solution of the steady state, we have \cite{SCBA-1}
\begin{align}\label{Arhost-Ad}
{\cal A}^{(\pm)}_{\alpha\mu\bar\rho}
&=\int^\infty_{-\infty}\frac{d\omega}{2\pi}\,
\Gamma^{(\pm)}_{\alpha\mu}(\omega)
{\cal U}(\pm\omega)[a^{\pm}_\mu\bar\rho].
\end{align}
Thus, after some calculations,
${\cal U}(\pm\omega)[a^{\pm}_\mu\bar\rho]$
can be expressed as
%\bsube
\begin{align}
{\cal U}(\omega)[a^{\dg}_\mu\bar\rho]
&=\left[\lambda^+_\mu(\omega)|\mu\ra\la 0|
+\kappa^+_\mu(\omega)(-1)^{\mu}|d\ra\la \bar \mu|\right] , \nonumber
\\
{\cal U}(-\omega)[a_\mu\bar\rho]&=\left[\lambda^-_\mu(\omega)|0\ra\la \mu|
+\kappa^-_\mu(\omega)(-1)^{\mu}|\bar \mu\ra\la d|\right] ,
\end{align}
%\esube
where
%\bsube
\begin{align}
\lambda^+_\mu(\omega)&=i\frac{ \Pi^{-1}_{1\mu}(\omega)\bar\rho_{00}
-\Sigma^-_{\bar\mu}(\omega)\bar\rho_{\bar\mu\bar\mu}}
{ \Pi^{-1}_{\mu}(\omega)\Pi^{-1}_{1\mu}(\omega)},  \nonumber
\\
\lambda^-_\mu(\omega)&=i\frac{ \Pi^{-1}_{1\mu}(\omega)\bar\rho_{\mu\mu}
-\Sigma^-_{\bar\mu}(\omega)\bar\rho_{dd} }
{ \Pi^{-1}_{\mu}(\omega)\Pi^{-1}_{1\mu}(\omega)}, \nonumber
\\
\kappa^+_\mu(\omega)&=i\frac{ -\Sigma^+_{\bar\mu}(\omega)\bar\rho_{00}
+\Pi^{-1}_{\mu}(\omega)\bar\rho_{\bar\mu\bar\mu} }
{ \Pi^{-1}_{\mu}(\omega)\Pi^{-1}_{1\mu}(\omega)},  \nonumber
\\
\kappa^-_\mu(\omega)&=i\frac{ -\Sigma^+_{\bar\mu}(\omega)\bar\rho_{\mu\mu}
+\Pi^{-1}_{\mu}(\omega)\bar\rho_{dd} }
{ \Pi^{-1}_{\mu}(\omega)\Pi^{-1}_{1\mu}(\omega)}. \nonumber
\end{align}
%\esube
Here, we introduced
$\Pi^{-1}_{ \mu}(\omega)=\omega-\epsilon_\mu-\Sigma_{0\mu}(\omega)
 -\Sigma^+_{\bar\mu}(\omega)$, and
$\Pi^{-1}_{1\mu}(\omega)=\omega-\epsilon_\mu-U-\Sigma_{0\mu}(\omega)
-\Sigma^-_{\bar\mu}(\omega)$.
The self-energies $\Sigma_{0\mu}(\omega)$ and $\Sigma^{\pm}_{\mu}(\omega)$
are given by
%\begin{subequations}\label{SEN}
\begin{align}
\Sigma_{0\mu}(\omega)
&=\int^\infty_{-\infty} \frac{d\omega'}{2\pi}
\frac{\Gamma_{\mu}(\omega')}{\omega-\omega' +i0^+} , \nl
%\end{align}
% \begin{align}
\Sigma^{\pm}_{\mu}(\omega)
&=\int^\infty_{-\infty} \frac{d\omega'}{2\pi}
\frac{\Gamma^{(\pm)}_{\mu}(\omega')}{\omega-\epsilon_{\bar\mu}
+\epsilon_\mu-\omega' +i0^+}
\nl&
+\int^\infty_{-\infty} \frac{d\omega'}{2\pi}
\frac{\Gamma^{(\pm)}_{\mu}(\omega')}{\omega-E_d+\omega' +i0^+}.
\end{align}
%\end{subequations}
Then, we find the solution of $\bm\varphi(\omega)$ as
\begin{align}\label{Kondo}
& \bm\varphi(\omega)
=
\frac{i(1-n_{\bar\mu})}
{\omega-\epsilon_\mu-\Sigma_{0\mu}
 +U\Sigma^+_{\bar\mu}(\omega-\epsilon_\mu-U-\Sigma_{0\mu}
 -\Sigma_{\bar\mu})^{-1}
 }   \nl
%\nla
& ~
+\frac{ i n_{\bar\mu}}
{\omega-\epsilon_\mu-U-\Sigma_{0\mu}
 -U\Sigma^-_{\bar\mu}(\omega-\epsilon_\mu-\Sigma_{0\mu}
 -\Sigma_{\bar\mu})^{-1}
 },
\end{align}
where $n_\mu=\rho_{\mu\mu}+\rho_{dd}$, and
$1-n_\mu=\rho_{\bar\mu\bar\mu}+\rho_{00}$.
This result coincides precisely with the one from the EOM
technique of the nonequilibrium Green's function \cite{Jau96},
which contains the remarkable nonequilibrium Kondo effect.

At high temperatures, \Eq{Kondo} reduces to
\begin{align}\label{HF}
\bm\varphi_{HF}(\omega) = \frac{i(1-n_{\bar\mu})}
{\omega-\epsilon_\mu-\Sigma_{0\mu} }
+  \frac{ i n_{\bar\mu}}
{\omega-\epsilon_\mu-U-\Sigma_{0\mu}  }.
\end{align}
Here, we use $\bm\varphi_{HF}$ to indicate the result
at the level of a mean-field Hatree-Fock approximation.
\Eq{HF} can also be derived from the EOM technique
at a lower-order cutoff \cite{Jau96}.
As a result, even the simple broadening effect
contained in \Eq{HF} goes beyond the scope of
the second-order ME.
In Fig.\ 13, we plot the current-voltage relation
based on \Eq{Kondo} against that from \Eq{HF}.

\begin{figure}[h]
\includegraphics[width=6cm]{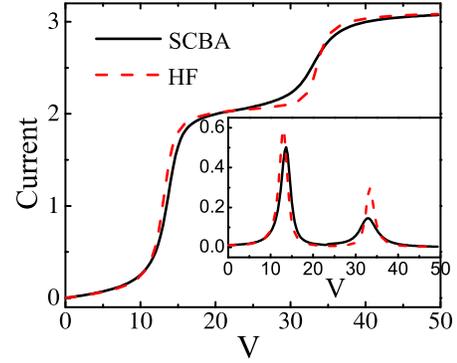}
\caption{
Coulomb staircase in the current-voltage curve.
Inset: the corresponding differential conductance.
The result based on \Eq{Kondo} is plotted
against the Hatree-Fock (HF) solution \Eq{HF}.
Parameters: $\Gamma_L=\Gamma_R=\Gamma/2$,
$\epsilon_0=7\Gamma$, $U=10\Gamma$, and $k_BT=0.1\Gamma$.
The bias voltage is set as $\mu_L=-\mu_R=eV/2$
by taking the zero-bias Fermi level as energy reference.  }
\end{figure}

\subsection{Formulation of the $n$-SCBA-ME}

Let us turn to the construction
of the $n$-resolved SCBA-ME ($n$-SCBA-ME),
along the same line of constructing the
$n$-resolved second-order ME \cite{SCBA-2,SCBA-3}.
The basic idea is to split the Hilbert space of the reservoirs
into a set of subspaces, each labeled by $n$.
Then, the average (trace) over each subspace is calculated,
and the corresponding {\it conditional}
reduced density matrix is defined as $\rho^{(n)}(t)$.
To be specific, consider the $\rho^{(n)}(t)$
conditioned on the electron number
arrived to the right lead, which obeys the following equation \cite{SCBA-2,SCBA-3}
\begin{align}\label{nME-scba}
\dot{\rho}^{(n)}
   & =  -i {\cal L}\rho^{(n)} -  \sum_{\mu}
%\nl&\quad
   \Big\{\big [a_{\mu}^{\dg} {\cal A}_{\mu\ti{\rho}^{(n)}}^{(-)}
 %\nl&\quad
   +a_{\mu} {\cal A}_{\mu\ti{\rho}^{(n)}}^{(+)}
 %\nl &\quad
    -  {\cal A}_{L\mu\ti{\rho}^{(n)}}^{(-)}a_{\mu}^{\dg}
  \nl
&  -  {\cal A}_{L\mu\ti{\rho}^{(n)}}^{(+)}a_{\mu}
 %\nl&\quad
   -  {\cal A}_{R\mu\ti{\rho}^{(n-1)}}^{(-)}a_{\mu}^{\dg}
 %\nl&\quad
    -  {\cal A}_{R\mu\ti{\rho}^{(n+1)}}^{(+)}a_{\mu}\big]
 %\nl&\qquad
 +{\rm H.c.} \Big\}  .
%\nl & \equiv -i {\cal L}\rho^{(n)} + {\cal R}_{\ti{\rho}^{(n)}}
%   + {\cal R}^{(+)}_{\ti{\rho}^{(n+1)}}
%   + {\cal R}^{(-)}_{\ti{\rho}^{(n-1)}}  .
\end{align}
Here, $ {\cal A}^{(\sigma)}_{\alpha\mu\ti{\rho}^{(n)}}(t)
=\sum_\nu\int^t_0 d\tau C^{(\sigma)}_{\alpha\mu\nu}(t-\tau)
[ \ti{\rho}_j^{(n)}(t,\tau)] $, while the summation over $\nu$
is appropriate in regard to the abbreviation $j=\{\nu,\sigma\}$.
In \Eq{nME-scba},
the appearance of $\tilde{\rho}^{(n\pm 1)}_j(t,\tau)$
is owing to a more tunneling event (forward/backword)
involved in the process of the corresponding terms.
In particular, $\ti{\rho}_j^{(n)}(t,\tau)$
is the $n$-dependent version of the quantity
$\ti{\rho}_j(t,\tau)={\cal U}(t,\tau)[a^{\sigma}_{\nu}\rho(\tau)]$,
satisfying the EOM according to \Eq{tirhoj}:
\begin{align}\label{nME-second}
\dot{\ti\rho}^{(n)}_{j}
   &=  -i {\cal L}\ti\rho^{(n)}_{j} -  \sum_{\mu}
%\nl&\quad
   \Big\{\big [a_{\mu}^{\dg} A_{\mu\tilde{\rho}^{(n)}_{j}}^{(-)}
 %\nl&\quad
   +a_{\mu} A_{\mu\tilde{\rho}^{(n)}_{j}}^{(+)}
 %\nl &\quad
    +  A_{L\mu\tilde{\rho}^{(n)}_{j}}^{(-)}a_{\mu}^{\dg}
\nl& %\quad
   +  A_{L\mu\tilde{\rho}^{(n)}_{j}}^{(+)}a_{\mu}
 %\nl&\quad
   +  A_{R\mu\tilde{\rho}^{(n-1)}_{j}}^{(-)}a_{\mu}^{\dg}
 %\nl&\quad
    +  A_{R\mu\tilde{\rho}^{(n+1)}_{j}}^{(+)}a_{\mu}\big]
 %\nl&\qquad
 +{\rm H.c.} \Big\} .
% \nl&
% \equiv -i {\cal L}\ti\rho^{(n)}_{j}
% + R_{\ti\rho^{(n)}_{j}} + R^{(+)}_{\ti\rho^{(n+1)}_{j}}
% + R^{(-)}_{\ti\rho^{(n-1)}_{j}}
\end{align}
In this equation, we introduced
$ A^{(\sigma')}_{\alpha\mu\ti\rho^{(n)}_{j}}(t)
 =\sum_{\nu'}\int^t_\tau dt' C^{(\sigma')}_{\alpha\mu\nu'}(t-t')
\left\{ e^{-i{\cal L}(t-t')}[a^{\sigma'}_{\nu'}
\tilde{\rho}_j^{(n)}(t')]\right\}$.

The $n$-resolved ME contains important information
and can be used in a wide variety of applications.
Next, we focus on calculating the shot noise spectrum
$S(\omega)$, using the MacDonald's formula
$S(\omega)=2\omega\int^{\infty}_{0}dt \sin(\omega t)
\frac{d}{dt}\langle n^{2}(t)\rangle$,
where
$\langle n^{2}(t)\rangle = \sum_{n}n^{2}P(n,t)= {\rm Tr}
\sum_{n} n^{2} \rho^{(n)}(t)$, and the $n$-counting
starts with the steady state ($\bar{\rho}$).
Based on \Eq{nME-scba}, one can express
$\frac{d}{dt} \langle n^{2}(t)\rangle$
in terms of ${\cal A}^{(\sigma)}_{R\mu\bar{\rho}}(t)$
and ${\cal A}^{(\sigma)}_{R\mu\ti{N}}(t)$.
The former has been introduced in \Eq{SCBA-ME},
needing only to replace $\rho(\tau)$ by $\bar{\rho}$.
The latter reads $ {\cal A}^{(\sigma)}_{R\mu\ti{N}}(t)
=\sum_\nu\int^t_0 d\tau C^{(\sigma)}_{R\mu\nu}(t-\tau)
[ \ti{N}_j(t,\tau)]$, where
$\ti{N}_j(t,\tau)=\sum_n n \ti{\rho}_j^{(n)}(t,\tau)$,
noting also the abbreviation $j=\{\nu,\sigma\}$
which makes the summation over $\nu$ reasonable.
Then, the MacDonald's formula becomes:
\begin{align}\label{Sw-scba}
S(\omega)&= 2\omega {\rm Im}\sum_{\mu}{\rm Tr}\Big\{ 2 \big[
 {\cal A}_{R\mu \ti{N}}^{(-)}(\omega)a_{\mu}^{\dg}
 %\nl&\quad
    - {\cal A}_{R\mu \ti{N}}^{(+)}(\omega)a_{\mu}\big]
 \nl&\qquad\qquad
  %\nl&\qquad
  +  \big[
  {\cal A}_{R\mu\bar{\rho}}^{(-)}(\omega)a_{\mu}^{\dg}
 %\nl&\quad
    + {\cal A}_{R\mu\bar{\rho}}^{(+)}(\omega)a_{\mu}\big]
  \Big\} .
\end{align}
This result is obtained after Laplace transforming
${\cal A}^{(\sigma)}_{R\mu\bar{\rho}}(t)$
and ${\cal A}^{(\sigma)}_{R\mu\ti{N}}(t)$. More explicitly,
\be
%\begin{align}\label{SW-1}
{\cal A}^{(\sigma)}_{R\mu\bar{\rho}}(\omega)
= \sum_\nu\int^\infty_{-\infty} \frac{d\omega'}{2\pi}
\Gamma^{(\sigma)}_{R\mu\nu}(\omega')
{\cal U}(\omega+\sigma\omega')[a^{\sigma}_\nu\bar{\rho}(\omega)], \nonumber
%\end{align}
\ee
where the Laplace transformation of the steady state
is given as $\bar{\rho}(\omega)=i\bar{\rho}/\omega$,
and the propagator ${\cal U}$ in frequency domain
is defined through \Eq{tirhoj}.
Another quantity,
${\cal A}^{(\sigma)}_{R\mu \ti{N}}(\omega)$ is given as
%\begin{align}\label{SW-2}
\be
{\cal A}^{(\sigma)}_{R\mu \ti{N}}(\omega)
= \sum_\nu\int^\infty_{-\infty} \frac{d\omega'}{2\pi}
\Gamma^{(\sigma)}_{R\mu\nu}(\omega')
\ti{\cal U}(\omega+\sigma\omega')[a^{\sigma}_{\nu}N(\omega)].  \nonumber
\ee
In deriving this result, we introduced an additional propagator
through $\ti{N}_j(t,\tau)=\ti{\cal U}(t-\tau)\ti{N}_j(\tau)$,
where $\ti{N}_{j}(\tau)=a^{\sigma}_{\nu} N(\tau)$ as the initial condition
which is defined by $N(\tau)=\sum_n n \rho^{(n)}(\tau)$.
$\ti{\cal U}(\omega)$ and $N(\omega)$ can be obtained via
Laplace transforming the following EOMs.
%%======================
{\it (i)} For $N(\omega)$,
based on the $n$-SCBA-ME we obtain:
\begin{align}\label{N-t}
  \dot{N}(t) &=-i{\cal L}N(t)
  - \sum_{\mu\sigma}\Big\{\big[a^{\bar\sigma}_\mu,
  {\cal A}^{(\sigma)}_{\mu N}(t)\big]
  +{\rm H.c.}  \Big\}  \nl
& ~~ + \sum_{\mu} \Big\{\big[ {\cal A}_{R\mu\bar{\rho}}^{(-)}a_{\mu}^{\dg}
    - {\cal A}_{R\mu\bar{\rho}}^{(+)}a_{\mu}\big] +{\rm H.c.} \Big\}  .
\end{align}
%=============================
{\it (ii)}
For $\ti{\cal U}(\omega)$, from \Eq{nME-second} we have
\begin{align}\label{Nj-second}
& \dot{\ti N}_{j}(t)
   =  -i {\cal L}\ti N_{j}(t)
   - \int^t_{\tau} dt' {\Sigma}^{(A)}_2(t-t')\ti N_{j}(t')
\nl& ~
 - \sum_{\mu} \Big\{\big [A_{R\mu\ti{\rho}_{j}}^{(-)}(t)a_{\mu}^{\dg}
-A_{R\mu\ti{\rho}_{j}}^{(+)} (t)a_{\mu}\big]
 %\nl&\qquad
 +{\rm H.c.} \Big\}.
\end{align}
The self-energy superoperator ${\Sigma}^{(A)}_2(t-t')$
is referred to \Eq{Acmt} for the definition and interepation/discussion.
Similarly, as introduced in \Eq{nME-second}, we defined here
$ A^{(\sigma')}_{R\mu\ti\rho_{j}}(t)
 =\sum_{\nu'}\int^t_\tau dt' C^{(\sigma')}_{R\mu\nu'}(t-t')
\left\{ e^{-i{\cal L}(t-t')}[a^{\sigma'}_{\nu'}
\tilde{\rho}_j(t')]\right\}$.

For the convenience of applications, we summarize
the solving procedures in a simpler way as follows.
First, solve ${\cal U}(\omega)$ from \Eq{tirhoj}
and obtain $\rho(\omega)$ from \Eq{SCBA-ME};
then, extract $\tilde{\cal U}(\omega)$ from \Eq{Nj-second}
and $N(\omega)$ from \Eq{N-t}.
Using
$\tilde{\cal U}(\omega)$ and $N(\omega)$, we can directly calculate the noise spectrum of \Eq{Sw-scba}.

\subsection{Noise Spectrum: Illustrative Examples}

\subsubsection{Noninteracting quantum dot }

Let us first consider the simplest setup of transport
through a single-level quantum dot.
In the absence of the magnetic field and Coulomb interaction,
the spin degree of freedom is irrelevant.
Then, the system Hamiltonian is given as $H_S = \epsilon_0 a^{\dg}a$,
and the states involved in the transport are $|0\ra$ and $|1\ra$,
corresponding to the empty and occupied dot states.
Applying the solving protocol outlined above,
the shot noise spectrum can be directly obtained,
as shown in Fig.\ 14 by the solid curve.
As a comparison, in Fig.\ 14, we plot also
the results from the second-order
non-Markovian (nMKV) and Markovian (MKV) ME,
respectively, by the dashed and dotted curves. 
The former is based on Ref.\ \cite{Jin2011},
while the later is from the following analytic result \cite{Li053}
%{Luo2007}
%... // .. {\bf (add two refs..)... Jin2011; Luo2007.. }
\begin{align}
S(\omega)=2\bar{I} \left(
\frac{\Gamma^2_L+\Gamma^2_R+\omega^2}{\Gamma^2+\omega^2} \right),
\end{align}
where $\Gamma=\Gamma_L+\Gamma_R$ is assumed.
$\bar{I}$ is the steady state current, in large bias limit which
is given as $\bar{I}=\Gamma_L\Gamma_R/\Gamma$, while here
we account for the finite bias effect based on the SCBA-ME approach.

We observe that, quantitatively, the result from the $n$-SCBA-ME
modifies that from the second-order nMKV-ME, while qualitatively both revealing
a staircase behavior at frequency
around $\omega_{\alpha 0}=|\mu_{\alpha}-\epsilon_0|$.
Mathematically, the origin of the staircase is from
the time-nonlocal memory effect.
Physically, this behavior is owing to the detection-energy ($\omega$)
assisted transmission resonance between the dot and leads,
which experiences a sharp change when crossing the Fermi levels.
In high frequency regime, the noise spectrum from the $n$-SCBA-ME
coincides with that from the second-order nMKV-ME, while the latter
is given in Ref.\ \cite{Jin2011}
by the high frequency limit
as $S(\omega\rightarrow\infty)=\Gamma_R$.
This directly leads to a Fano factor
as $F=S/2\bar{I}=(1+\Gamma_R/\Gamma_L)/2$.
Therefore, it can be Poissonian, sub-Poissonian,
and super-Poissonian, depending on the symmetry factor $\Gamma_R/\Gamma_L$.
In contrast, the second-order nMKV-ME predicts
a Poissonian result, $F(\omega\rightarrow\infty)=1$.

We would like to point out that the second-order MKV-ME is only applicable
in the low frequency regime of
$\omega<\omega_{\alpha 0}=|\mu_{\alpha}-\epsilon_0|$.
This is in consistency with the fact that
the high frequency regime corresponds to a short
timescale where the non-Markovian effect is strong,
while the low frequency regime corresponds to a
long timescale where the non-Markovian effect diminishes.

\begin{figure}
\includegraphics[scale=0.25]{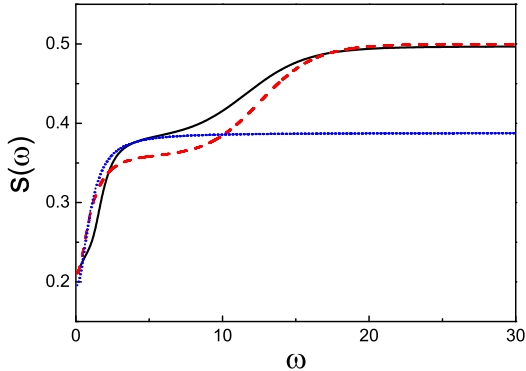}
\caption{
Shot noise spectrum through a single-level noninteracting quantum dot,
from the SCBA-ME (solid curve), the 2nd-nMKV-ME (dashed curve)
and the 2nd-MKV-ME (dotted curve), respectively.
Parameters: $\Gamma_L=\Gamma_R=0.5$,
$\mu_L=-\mu_R=7.5$,
$\epsilon_0=5$, $k_BT=2$ and $W=100$.  }
\end{figure}

\subsubsection{Coulomb-Blockade quantum dot }

This is the system described by \Eq{Ands-H}.
Here we first consider
the noise spectrum in the Coulomb-Blockade (CB) regime,
and leave the Kondo regime to next subsection. 
The CB regime of single occupation is characterized by
$\epsilon_0+U>\mu_L>\epsilon_0>\mu_R$.
For the purpose of comparison, we quote
the result from the second-order MKV-ME \cite{Li053}
% {Luo2007}(..Luo2007..)
\begin{align}\label{CBSW}
S(\omega) = 2\bar{I} \left[
\frac{4\Gamma^2_L+\Gamma^2_R+\omega^2}
{(2\Gamma_L+\Gamma_R)^2+\omega^2} \right].
\end{align}
In large bias limit, i.e., the Fermi levels far from
$\epsilon_0$ and $\epsilon_0+U$, the steady state current
is given as $\bar{I}=2\Gamma_L\Gamma_R/(2\Gamma_L+\Gamma_R)$.
However, in numerical simulation, we account for the finite bias effect by
inserting the steady state current from the SCBA-ME approach into \Eq{CBSW}.
For obtaining \Eq{CBSW},
the double occupancy of the dot is excluded
because the energy is out of the bias window.
In the $n$-SCBA-ME treatment, however,
all the four basis states should be included.

In Fig.\ 15, we display the main result of the noise spectrum in the CB regime,
where a couple of non-Markovian resonance steps are revealed at frequencies
around $\omega_{\alpha 0}=|\mu_{\alpha}-\epsilon_0|$
and $\omega_{\alpha 1}=\epsilon_0+U-\mu_{\alpha}$.
We find that the resonance steps in high frequency regime
are enhanced by the Coulomb interaction, while the low frequency
spectrum has remarkable ``renormalization" effect compared to \Eq{CBSW}.
In addition to the result under the wide band limit (WBL),
in Fig.\ 15, we also show the bandwidth effect by two more curves.
We see that, for finite-bandwidth leads, the noise spectrum
diminishes at the high frequency limit.
This is because the energy ($\omega$) absorption/emission
of detection restricts the channels for electron transfer
between the dots and leads.

\begin{figure}
\includegraphics[scale=0.25]{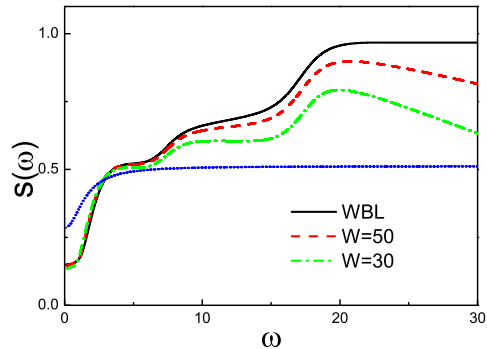}
\caption{
Shot noise spectrum through an interacting quantum dot
in a Coulomb blockade regime defined as
$\epsilon_0+U>\mu_L>\epsilon_0>\mu_R$.
Results in the wide band limit (WBL)
and for the finite bandwidths are shown
in comparison with the one from the 2nd-MKV-ME (dotted line).
Parameters: $\Gamma_L=\Gamma_R=0.5$, $\mu_L=-\mu_R=5$,
$\epsilon_{\uparrow}=\epsilon_{\downarrow}=\epsilon_0=2$,
$U=10$, and $k_BT=2$. We find staircases
appearing at $\omega=\epsilon_0-\mu_R=7$
and $\epsilon_0+U-\mu_R=17$.  }
\end{figure}

%%\newpage

\subsubsection{Nonequilibrium Kondo dot}

The nonequilibroum Kondo system, with the Anderson impurity model
realized by transport through a small quantum dot,
has attracted intensive attentions
\cite{Gor98,Kou98,Gla04,Ng88,Her91,MW92,MW9193,Ra94,Mar06,Gor08,Yan12}.
Compared to the equilibrium Kondo effect,
the nonequilibrium is characterized by
a finite chemical potential difference of the two leads.
As a result, the peak of the density of states (spectral function)
splits into two peaks pinned at each chemical potential.
The two peak structure is difficult to probe directly,
by the usual dc measurements.
Nevertheless, the shot noise might be a promising quantity
to reveal the nonequilibrium Kondo effect,
although much less is known about it.
Despite the low-frequency noise measurements \cite{De09,Hei08},
so far there are no reports
on the finite-frequency (FF) noise measurements.
However, several theoretical studies \cite{Ng97,Her98,Kon07,Moc11}
revealed diverse signatures (Kondo anomalies) in the FF noise spectra,
such as an ``upturn" \cite{Ng97} or a spectral ``dip" \cite{Moc11}
appeared at frequencies $\pm eV/\hbar$ ($V$ is the bias voltage),
as well as the Kondo singularity (discontinuous slope)
at frequencies $\pm 2eV/\hbar$ in Ref.\ \cite{Her98},
or at $\pm eV/2\hbar$ in Ref.\ \cite{Moc11}.
In addition, it was noted in Ref.\ \cite{Her98} that
the minimum (dip) developed at $\pm eV/\hbar$
is not relevant to the Kondo effect,
since in the noninteracting case the noise
has a similar discontinuous slope at $\pm eV/\hbar$ as well \cite{Her98}.

The system Hamiltonian is the same as \Eq{Ands-H}.
Following the solving protocol outlined above, we obtain the
noise spectrum in the Kondo regime, as shown in Fig.\ 16.
Remarkably, we observe a profound ``dip" behavior (Kondo signature)
in the noise spectrum at frequencies $\omega=\pm V/2$,
as particularly demonstrated by a couple of voltages.
We attribute this behavior to the emergence of the
Kondo resonance levels (KRLs) at the Fermi surfaces,
i.e., at $\mu_L=V/2$ and $\mu_R=-V/2$.
In steady state transport, it is well known that
the KRLs are clearly reflected in the spectral function.
In the ME,
the KRLs structure is hidden in the self-energy terms,
which characterize the tunneling process
and define the transport current.
Similarly, the noise spectrum is affected,
particularly in the Kondo regime,
by the self-energy process in frequency domain
based on the same ME.
This explains the emergence of the spectral dip appearing
at the same KRLs (i.e., at $\omega=\pm V/2$).

Alternatively, as a heuristic picture, one may imagine to include
the KRLs as basis states in propagating $\rho(t)$,
which is implied in the current correlation function.
In a usual case, when the level spacing is larger than the broadening,
the diagonal elements of the density matrix decouple to
the evolution of the off-diagonal elements.
However, in the Kondo system, the diagonal and off-diagonal elements
are coupled to each other, through the complicated self-energy processes.
This feature would bring the coherence evolution
described by the off-diagonal elements, with characteristic
energies of the KRLs and their difference, into the diagonal elements
which contribute directly to the the second current measurement
in the correlation function $\la I(t) I(0) \ra$.
Then, one may expect three coherence energies, $\pm V/2$ and $V$,
to participate in the noise spectrum.
Indeed, the dip emerged in Fig.\ 16 reveals the {\bf coherence-induced}
oscillation at the frequencies $\pm V/2$,
while the other one at the higher frequency $V$
(observed in by Moca {\it et al.}\cite{Moc11} in the case of infinite $U$)
is smeared in our finite $U$ system by the rising noise with frequency.

\begin{figure}
\includegraphics[scale=0.25]{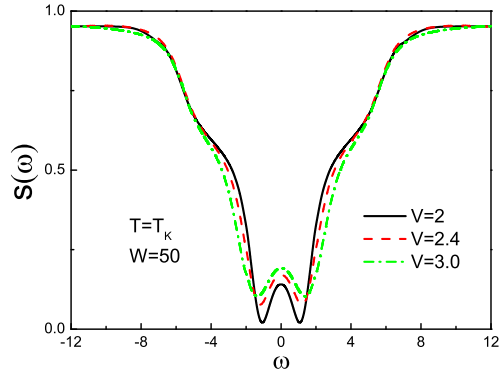}
\caption{
Shot noise spectrum in the Kondo regime,
for several bias voltages ($\mu_L=-\mu_R=V/2$).
Parameters: $\Gamma_L=\Gamma_R=\Gamma_0=0.5$,
$\epsilon_{\uparrow}=\epsilon_{\downarrow}=\epsilon_0=-2$, and $U=6$.
The Kondo temperature is determined by
$T_K=\frac{U}{2\pi}\sqrt{\frac{-2U\Gamma_0}{\epsilon_0(U+\epsilon_0)}}
\exp[\frac{\pi\epsilon_0(U+\epsilon_0)}{2U\Gamma_0}]$,
from which we obtain $T_{K}=0.144$.    }
\end{figure}

% =========================================================
\section{Concluding Remarks}

{\flushleft In summary}, 
we have reviewed the formulation of
particle-number($n$)-resolved master equation ($n$-ME) approach,
and its application to quantum measurement
and quantum transport in mesoscopic devices.
The formalism under the (standard) second-order Born approximation
is particularly simple and can be reliably applied
to many practical problems.
Importantly, the $n$-ME version is extremely appropriate
for studying the shot noise and counting statistics
(including also the large-deviation analysis),
which encode additional dynamic information
beyond the stationary current.
The convenient application of the $n$-ME approach
was illustrated by a couple of examples.

However, the second-order Born approximation
does not fully capture the tunneling induced level-broadening
and other multiple forward-backward-process induced
correlation effects.
Therefore, the first limitation is that
the second-order ME to quantum transport
is valid only for large bias voltage.
On the other hand,  for interacting systems
(e.g., quantum dot or Anderson impurity),
the second-order ME cannot describe the cotunneling
and the nonequilibrium Kondo effect owing to the same reason.
To overcome these limitations, higher-order expansions
of the tunneling Hamiltonian are required, as performed
by a variety of cases in literature
\cite{Sch94,Sch96,Sch06,Yan080911,Yan2012,Wac05+10,CS11,
Leeu09,Galp09,Galp10,KG13}.

We have therefore reviewed a newly proposed ME
approach (and its $n$-resolved version) termed as SCBA-ME
(under the self-consistent Born approximation
when expanding the tunneling Hamiltonian).
The basic idea is replacing the free (system only) Green's function
in the second-order self-energy operator by an {\it effective} propagator
defined by the second-order ME.
We found that the effect of this
simple improvement is remarkable: it can recover not only
the exact result of (any) noninteracting transport
under {\it arbitrary} bias voltage
but also describe the cotunneling and nonequilibrium
Kondo effect in Coulomb interacting systems.

Finally, we mention that a similar idea
of modifying the free propagator in the tunneling
self-energy diagram by a dressed one was implemented
also in a couple of recent studies \cite{Galp09,Galp10,KG13}.
In Ref.\ \cite{KG13}, the specific Anderson impurity model was
solved heavily based on a diagrammatic technique,
but lacking a general formulation of basis-free master equation.
In Refs.\ \cite{Galp09,Galp10},
owing to inappropriately treating the dressed propagator
as a Markovian-Redfield generator, problems occurred
as mentioned in the concluding remarks in Ref.\ \cite{Galp10}:
`` $\cdots$ However, note that many important effects due to strong correlation
between the molecule and contacts observed at low temperatures
(e.g., Kondo) cannot be reproduced within our scheme.
We find that our scheme becomes unreliable in the region of the parameters
where coherences in the system eigenbasis
(i.e., coherences introduced through nondiagonal elements of
molecule-contact coupling matrix $\Gamma$) are bigger than
the inter-level separation and on the order of the diagonal elements
of the molecule-contact coupling matrix $\Gamma$".
Remarkably, the SCBA-ME approach reviewed in this work
overcomes all these drawbacks.

% =====================================================
\vspace{0.2cm}
{\flushleft \it Acknowledgments.}---
The author is grateful to many former students and collaborators
whose invaluable contributions constitute the main elements
of this review article. Some of them are:
Jinshuang Jin, Junyan Luo, Shikuan Wang, Hujun Jiao,
Yonggang Yang, Jun Li, Feng Li, Yu Liu, Jing Ping, Ping Cui,
Wenkai Zhang, Jiushu Shao, YiJing Yan, and Shmuel Gurvitz.
This work was supported by the NNSF of China
under No.\ 91321106 and the State ``973" Project
under Nos.\ 2011CB808502 \& 2012CB932704.

%% \clearpage

%\end{CJK*}
\end{document}